\documentclass{aastex61}

\usepackage{amsmath}

\received{}
\revised{}
\accepted{}

\shorttitle{Heating by driven kink and Alfv{\'e}n waves in coronal loops}
\shortauthors{Guo et al.}

\begin{document}

\title{Heating effects from driven transverse and Alfv{\'e}n waves in coronal loops}

\correspondingauthor{Mingzhe Guo}
\email{mingzhe.guo@kuleuven.be}

\author{Mingzhe Guo}
\affiliation{Institute of Space Sciences, Shandong University, Weihai 264209, China}
\affiliation{Centre for mathematical Plasma Astrophysics, Department of Mathematics, KU Leuven, 3001 Leuven, Belgium}

\author{Tom Van Doorsselaere}
\affiliation{Centre for mathematical Plasma Astrophysics, Department of Mathematics, KU Leuven, 3001 Leuven, Belgium}

\author{Kostas Karampelas}
\affiliation{Centre for mathematical Plasma Astrophysics, Department of Mathematics, KU Leuven, 3001 Leuven, Belgium}

\author{Bo Li}
\affiliation{Institute of Space Sciences, Shandong University, Weihai 264209, China}

\author{Patrick Antolin}
\affiliation{School of Mathematics and Statistics, University of St. Andrews, St. Andrews, Fife KY16 9SS, UK}

\author{Ineke De Moortel}
\affiliation{School of Mathematics and Statistics, University of St. Andrews, St. Andrews, Fife KY16 9SS, UK}




\begin{abstract}

Recent numerical studies revealed that transverse motions of coronal loops can induce the Kelvin-Helmholtz Instability (KHI). 
This process could be important in coronal heating because it leads to dissipation of energy at small spatial-scale plasma interactions.
Meanwhile, 
small amplitude decayless oscillations in coronal loops have been discovered recently in observations of SDO/AIA. 
We model such oscillations in coronal loops and study wave heating effects, 
considering a kink and Alfv{\'e}n driver separately and a mixed driver at the bottom of flux tubes.
Both the transverse and Alfv{\'e}n oscillations can lead to the KHI.
Meanwhile, 
the Alfv{\'e}n oscillations established in loops will experience phase mixing.
Both processes will generate small spatial-scale structures, 
which can help the dissipation of wave energy. 
Indeed, 
we observe the increase of internal energy and temperature in loop regions.
The heating is more pronounced for the simulation containing the mixed kink and Alfv{\'e}n driver.
This means that the mixed wave modes can lead to a more efficient energy dissipation in the turbulent state of the plasma and that the KHI eddies act as an agent to dissipate energy in other wave modes. 
Furthermore, 
we also obtained forward modelling results using the FoMo code. 
We obtained forward models which are very similar to the observations of decayless oscillations. 
Due to the limited resolution of instruments, 
neither Alfv{\'e}n modes nor the fine structures are observable. 
Therefore, 
this numerical study shows that Alfv{\'e}n modes probably can co-exist with kink modes, 
leading to enhanced heating.

\end{abstract}

\section{INTRODUCTION} 
\label{sec_intro}

A rich variety of MHD oscillations and waves have been observed in the highly structured solar atmosphere \citep[for recent reviews, e.g.,][]{2005LRSP....2....3N,2012RSPTA.370.3193D,2016SSRv..200...75N, 2015SSRv..190..103J}.
They may play an important role in coronal heating because of their capability of carrying energy \citep[e.g.,][]{2009SSRv..149..229T,2012RSPTA.370.3217P}.
In fact, 
analytical studies to reveal the various wave properties in magnetic structures date back to 1970s \citep{1975IGAFS..37....3Z,1979A&A....76...20W}.
In 1999, the first imaging observation of kink waves in active region loops was obtained by the Transition Region and Coronal Explorer \citep[TRACE,][]{1999ApJ...520..880A, 1999SoPh..187..261S, 1999Sci...285..862N}.
Since then, 
a large number of transverse waves were discovered in the solar atmosphere by modern instruments \citep[e.g.,][]{2005LRSP....2....3N,2005A&A...430L..65V, 2006RSPTA.364..417A,2009ApJ...697.1384T},
not only in coronal loops \citep[see][for a recent review]{2009SSRv..149..199R}, 
but in chromospheric spicules and mottles \citep[e.g.,][]{2007Sci...318.1574D,2009ApJ...705L.217H,2012NatCo...3E1315M,2012ApJ...750...51K,2013ApJ...779...82K}, 
large prominences \citep[e.g.,][]{2012LRSP....9....2A}, polar plumes \citep[e.g.,][]{2010ApJ...718...11G}, 
and coronal streamers \citep[e.g.,][]{2010ApJ...714..644C,2011ApJ...728..147C,2013ApJ...766...55K}.

The observed large-amplitude transverse oscillations generally undergo rapid damping in a couple of periods \citep{1999Sci...285..862N,2002SoPh..206...99A,2016A&A...585A.137G}.
Such damping is usually attributed to resonant absorption \citep{1988JGR....93.5423H,1992SoPh..138..233G,2002A&A...394L..39G} or phase mixing \citep{2015ApJ...803...43S}. 
The kink oscillations transfer into local Alfv{\'e}n modes when the kink frequency matches the local Alfv{\'e}n frequency,
therefore the transverse motion has an apparent decay.
This process is usually expected to happen in an inhomogeneous layer near the loop boundary.
On the other hand,
\citet{2015ApJ...803...43S} considered the phase mixing process simultaneously,
which is responsible for the wave energy transfer from large spatial-scale structures to small scale plasma interactions.
The real dissipation of wave energy at such small structures relies on resistivity and viscosity
 \citep{1994ApJ...421..360O,1998ApJ...493..474O}.

Recent simulations of transverse waves in coronal loops revealed that the Kelvin-Helmholtz instability occurs near the boundary of loops \citep{2008ApJ...687L.115T,2014ApJ...787L..22A,2017ApJ...836..219A, 2015A&A...582A.117M, 2017A&A...604A.130K,2017A&A...602A..74H,2017A&A...607A..77H}. 
The instability is generated due to the strong shear motions near the loop edges.
Meanwhile, the generation of azimuthal Alfv{\'e}n waves at resonant layers increases the velocity shear with the external plasma,
which can also enhance the instability and make the systems more unstable.
\citet{2014ApJ...787L..22A} revealed that even a small-amplitude ($\sim 3 {\rm km~s^{-1}}$) kink oscillation can lead to such an instability.
The importance of this Transverse Wave Induced Kelvin-Helmholtz (TWIKH) instability \citep{2014ApJ...787L..22A,2017ApJ...836..219A} is that it generates turbulent small structures.
This makes the wave energy dissipate much more easily in the small scale structures in the presence of transport coefficients or kinetic effects.
This is probably a crucial process for coronal heating \citep{2017A&A...604A.130K,2017A&A...602A..74H}.

Due to their incompressibility, 
Alfv{\'e}n waves are not easily detected by imaging instruments in the solar atmosphere.
 Their torsional motions would cause spectral line broadening,
making them detectable to spectrographs \citep{2003A&A...399L..15Z}.
\citet{2009Sci...323.1582J} reported the torsional Alfv{\'e}n waves in the chromosphere,
using the {\rm H}-$\alpha$ line in Solar Optical Universal Polarimeter (SOUP) of SST.
However,
the corresponding coronal observations remain unclear.
Theoretically, 
Alfv{\'e}n waves can be easily generated from the lower atmosphere \citep{1994A&A...283..232M,1999ApJ...526..478B}. 
\citet{2008A&A...478..553V} claimed that the wave energy flux through the photosphere becomes orders of magnitude smaller when considering the effects of partial ionization and collisions.
However,
the fast waves transfer their energy to upgoing Alfv{\'e}n waves in the conversion region.
The process is analogous to the resonant absorption mentioned above,
making the Alfv{\'e}n flux increase significantly in the lower atmosphere \citep{2012ApJ...746...68K,2013MNRAS.435.2589C,2018NatPh..14..480G}.
Similar to kink waves, the energy dissipation is an important issue.
\citet{1983A&A...117..220H} claimed that phase mixing occurs between different magnetic surfaces 
when Alfv{\'e}n waves propagate in the non-uniform magnetic structures.
Recent work by \citet{2018A&A...616A.125P} found that heating from phase mixing of Alfv{\'e}n modes in coronal loops with multi-harmonic oscillations is small.
However,
the Kelvin-Helmholtz instability can be induced for standing modes eventually due to the strong, localised velocity shear.
Such small turbulent structures induced through the instability can help wave energy dissipate more easily.

Recently, low-amplitude decayless transverse oscillations have been detected \citep{2013A&A...552A..57N,2013A&A...560A.107A,2015A&A...583A.136A}.
Imaging observations (such as SDO/AIA) revealed that this kind of apparently undamped oscillations is a common phenomenon in coronal loops with small average amplitudes.
\citet{2013A&A...552A..57N} initially interpreted such undamped regimes as the response of loops to external continuous drivers.
\citet{2016ApJ...830L..22A} explained such oscillations as combined effects of periodic brightening of TWIKH rolls 
and the limited resolution of instruments. 
\citet{2016A&A...591L...5N} proposed that these decayless oscillations were caused by interaction of loops with quasi-steady flows as self-oscillations.
Very recently, \citet{2017A&A...604A.130K,2018A&A...610L...9K} simulated such decayless transverse oscillations as coronal loops driven by transverse motions.

In this article, 
we aim to simulate such driven oscillations and study the heating effects,
considering a mixed kink and Alfv{\'e}n driver at one footpoint of a loop.
\citet{2010ApJ...710.1857M} claimed that turbulent photospheric motions can be observed by Hinode/SOT,
therefore it is reasonable to expect mixed motions at the footpoints of loops.
The mixed processes of KHI,
resonant absorption and phase mixing will greatly influence the heating effects.
Meanwhile, 
the decayless oscillations are ubiquitous in coronal loops,
so it is worthwhile to reveal their relation to coronal heating.
For comparison, 
pure Alfv{\'e}n and kink driver models are also considered respectively.
This manuscript is organized as follows.
Section \ref{sec_models} presents our basic setup of numerical models.
Apparent dynamics of the loops are presented in section \ref{sec_general_results}.
In section \ref{sec_energy}, 
we analyse the energy variations of the three models to examine the heating effects.
Forward modelling is performed in order to compare to real observations in section \ref{sec_fomo}.
Finally, section \ref{sec_conclusion} closes this paper with discussions and conclusions.

\section{NUMERICAL MODELS}
\label{sec_models}

\subsection{Equilibrium and Drivers}
\label{sec_sub_ini}

We consider three 3-D numerical models in our simulations.
They are all based on the same straight density enhanced magnetic tube, 
which is embedded in a uniform background plasma. 
We aim to model a coronal loop with a uniform magnetic field directed along the $z$-direction.
Similar models have been used in previous works \citep[e.g.,][]{2014ApJ...787L..22A,2017ApJ...836..219A, 2015A&A...582A.117M, 2017A&A...604A.130K, 2018A&A...610L...9K}.
The loop has an initial density ratio of $\rho_{\mathrm i}/\rho_{\mathrm e}=3$ 
(index i (e) denotes internal (external) values) and we consider a density profile given by 

\begin{equation}
\rho(x,y)=\rho_{\rm e}+(\rho_{\mathrm i}-\rho_{\mathrm e})\phi(x,y),
\label{eq_rho1}
\end{equation}
\begin{equation}
\phi(x,y)=\displaystyle \frac{1}{2} \left\{1-\tanh \left[b(\sqrt{x^2+y^2}/R-1)\right] \right\},
\label{eq_rho2}
\end{equation}
where, $x,y$ denote the coordinates in the plane perpendicular to the direction of the loop, 
which is fixed as the $z$-direction.
 $b$ sets the width of the boundary layer.
We choose $b=8$, which gives the width of the inhomogeneous layer $l\approx0.4R$,
 corresponding to a typical value estimated in coronal loops \citep{2002A&A...394L..39G}.
The initial parameters of the loop are shown in Table 1.
The loop length ($L=150{\rm Mm}$) and radius ($R=1{\rm Mm}$) are chosen within the range of observations \citep{1999Sci...285..862N,2002SoPh..206...99A}.
The density ratio here is chosen according to the estimated value in \citet{2003ApJ...598.1375A}.

We consider a uniform temperature loop ($T_{\rm i}=T_{\rm e}=1{\rm MK}$), 
so the average temperature increase due to the mixing between the colder tube and hotter background corona is avoided \citep{2017A&A...604A.130K}. 
Therefore it will be easier to identify the true wave heating effects. 
To maintain the magnetostatic pressure balance, 
the magnetic field has a slight variation from internal $B_ {\rm i}=50 {\rm G}$ to external $B_ {\rm e}=50.07 {\rm G}$.

 The magnetic field (50${\mathrm G}$) here is larger than previous models \citep[e.g.,][]{2014ApJ...787L..22A,2017ApJ...836..219A,2017A&A...604A.130K} and observations \citep[e.g.,][]{2001A&A...372L..53N,2007A&A...473..959V,2008A&A...487L..17V,2016NatPh..12..179J}.
In this case,
the energy input into the models is increased in order to obtain more noticeable heating effects.

\begin{table}[tbp]
  \begin{center}
  \caption{Parameters used in simulations} \label{tab_parameters}
 \begin{tabular}{ccccccc}
\\
\hline
\hline
  &  \multicolumn2c{Parameters}  &   \multicolumn2c{ } &   \multicolumn2c{Values}  \\
\hline
        & \multicolumn2c{Loop length $L$ (Mm)}   &   \multicolumn2c{ }   & \multicolumn2c{150}   \\
         & \multicolumn2c{Loop radius $R$ (Mm)}  &   \multicolumn2c{ }   & \multicolumn2c{1}  \\
         & \multicolumn2c{Internal density $\rho_{\rm i}$ (${\mathrm {g~cm^{-3}}}$)} & \multicolumn2c{ } & \multicolumn2c{$2.5\times10^{-15}$}  \\
         & \multicolumn2c{Density ratio $\rho_{\rm i}/ \rho_{\rm e}$}   &   \multicolumn2c{ }  & \multicolumn2c{3}  \\
         & \multicolumn2c{Temperature $T$ (MK)}  &   \multicolumn2c{ }   & \multicolumn2c{1 }  \\
         & \multicolumn2c{Magnetic field $B_{\rm i}$ (G)}  &   \multicolumn2c{ }   & \multicolumn2c{50}  \\
        \hline
\hline
\end{tabular}
\end{center}
\end{table}

In order to investigate the heating effects of different wave modes, 
we employ three models with the same initial parameters in Table 1,
but different drivers on the bottom footpoint ($z=0$). 
The first driver is a continuous, mono-periodic ``dipole-like'' driver, 
which is similar to \citet{2010ApJ...711..990P} and \citet{2017A&A...604A.130K}.
The time-dependent velocity inside the loop ($r< R$) is 
\begin{equation}
\mathbf{v}_{\rm i} =v_0\left[\sin\left(\frac{2\pi t}{P_{\rm k}}\right),0,0\right],
\label{eq_vi}
\end{equation}
where $v_0=2{\rm km~s^{-1}}$ is the amplitude of the velocity.
The period of the driver $P_{\rm k} = 87{\rm s}$, 
which corresponds to the predicted value for the fundamental kink mode \citep{1983SoPh...88..179E}.
The spatial dependence of the driver velocity outside the loop has the form 
\begin{equation}
\mathbf{v}_{\rm e} =v_0 R^2 \sin\left(\frac{2\pi t}{P_{\rm k}}\right)\left[ \frac{x^2-y^2}{(x^2+y^2)^2},\frac{2xy}{(x^2+y^2)^2},0 \right].
\label{eq_ve}
\end{equation}

We also use a transition layer between these two regions to avoid the numerical problems,
as \citet{2010ApJ...711..990P} and \citet{2017A&A...604A.130K} did. 
The profile is similar to the density profile given by Eq. (\ref{eq_rho1}) and Eq. (\ref{eq_rho2}).

The second driver is a broad band time-dependent torsional motion, 
mimicking Alfv{\'e}n oscillations inside a loop. 
The torsional driver is inspired by the one used in \citet{1999ApJ...526..478B}. 
To launch this driver, $v_\theta$ is described as 

\begin{eqnarray}
v_\theta=v_0 \sin\left(\displaystyle\frac{2\pi t}{P_{\rm A}(r)}\right)
\begin{cases}
\left(\displaystyle\frac{2r}{R}\right)^2 \left(\displaystyle\frac{2r}{R}-2\right)^2, & r/R\leq1 \\
 0,                                                 & r/R>1
\end{cases}
\label{eq_vtheta}
\end{eqnarray}
where $v_0$ keeps the same value as the one of the kink driver. 
The period $P_{\rm A}$ is a function of radial distance because we have a non-uniform transverse density distribution. 
It is given by $P_{\rm A}(r) = 2L/v_{\rm A}(r) = 2L\sqrt{\mu_0 \rho(r)}/B(r)$, 
varying from its internal value of $106 {\rm s}$ to the loop boundary ($r=R$) value of $87 {\rm s}$. 
Using these periods,
we can establish Alfv{\'e}n oscillations in the uniform region ($r<0.8R$) and the inhomogeneous region ($0.8R \leq r\leq R$) with the corresponding periods on the different magnetic surfaces.

Finally,
the third driver is a mixed Alfv{\'e}n and kink driver. 
 We consider both transverse velocity (given by Eq. (\ref{eq_vi}) and Eq. (\ref{eq_ve})) and torsional motions (given by Eq. (\ref{eq_vtheta})) simultaneously during the entire simulation. 
 Therefore,
 the energy provided by the mixed driver, i.e. input energy, is at the same level as the sum of the other two drivers. 
  
For simplicity, hereafter we name the kink driver model as ``K-model", 
the Alfv{\'e}n driver model as ``A-model'' and the mixed driver model as ``M-model".
In our K-model and M-model, 
the drivers follow the motions of loops, 
making sure that the internal loop regions will always have a uniform velocity.

\subsection{Numerical setup}
\label{sec_sub_setup}

To solve the 3-D time-dependent MHD equations, 
we use the PLUTO code \citep{2007ApJS..170..228M}.
A second-order parabolic spatial scheme is used for integration, 
the numerical fluxes are computed by a Roe Riemann solver.
Meanwhile, 
a third-order Runge-Kutta algorithm is used to advance the solution to the next time level. 
The simulation domain is ${\rm \left[-8,8\right] Mm \times \left[-8,8\right] Mm \times \left[0,150\right] Mm}$.
To resolve the motions of the drivers near the footpoints, 
we adopt a stretched mesh with 5 cells from 0 to $R$ and
a uniform grid of 95 points from $R$ to $L$ in the $z$-direction,.
For the $x$ and $y$ directions, 256 non-uniformly spaced cells are adopted, respectively.  
The resolution is up to ${\rm 20~km}$ in the region of $|x,y| \leq 2 {\rm Mm}$.
The following simulations show that this resolution is high enough to observe small structures induced by waves and instabilities.

In order to establish standing waves in loops,
we fix the velocities at $z=L$ to be zero to mimic loops anchored in the photosphere. 
The other variables there are set to obey Neumann-type (zero-gradient) conditions. 
The $z$-component velocities at the bottom footpoint ($z=0$) are antisymmetric and $v_x,v_y$ are described by the drivers.
 All the lateral boundaries are set to be outflow conditions.  

\section{GENERAL NUMERICAL RESULTS}
\label{sec_general_results}

We ran simulations until $t=1500 {\rm s}$ for all three models,
corresponding to roughly 14-17 periods.
The maximum displacements the loops experienced are less than 1Mm, 
to allow us to concentrate on the subdomain of $|x,y|\leq 2{\rm Mm},0\leq z\leq 150{\rm Mm}$, 
which is the domain with the highest resolution in the $x,y$ directions.

\subsection{KHI eddies, resonant absorption and phase mixing}
\label{sec_sub_eddies}

\begin{figure*}
\gridline{\fig{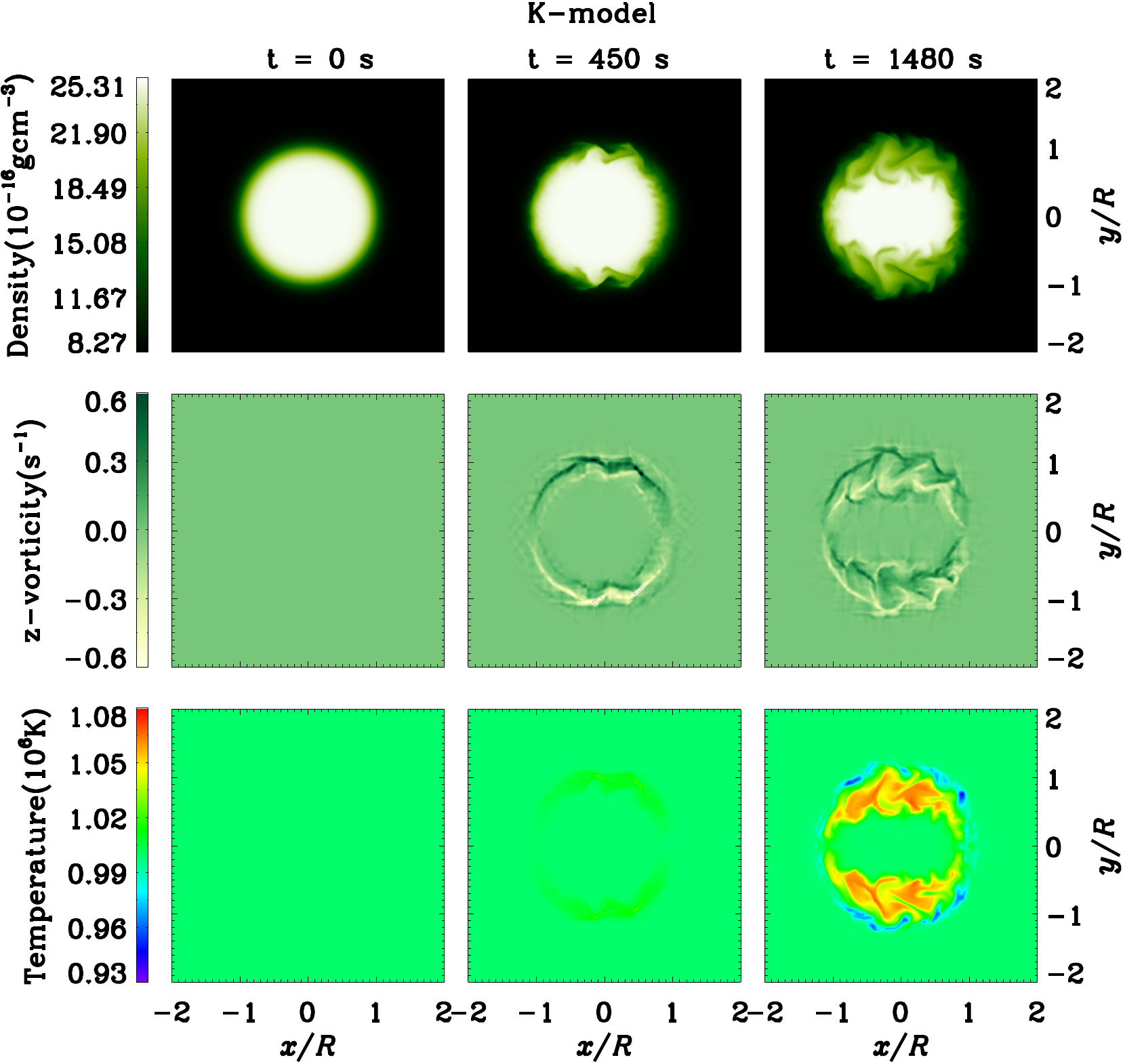}{0.33\textwidth}{(a)}
          \fig{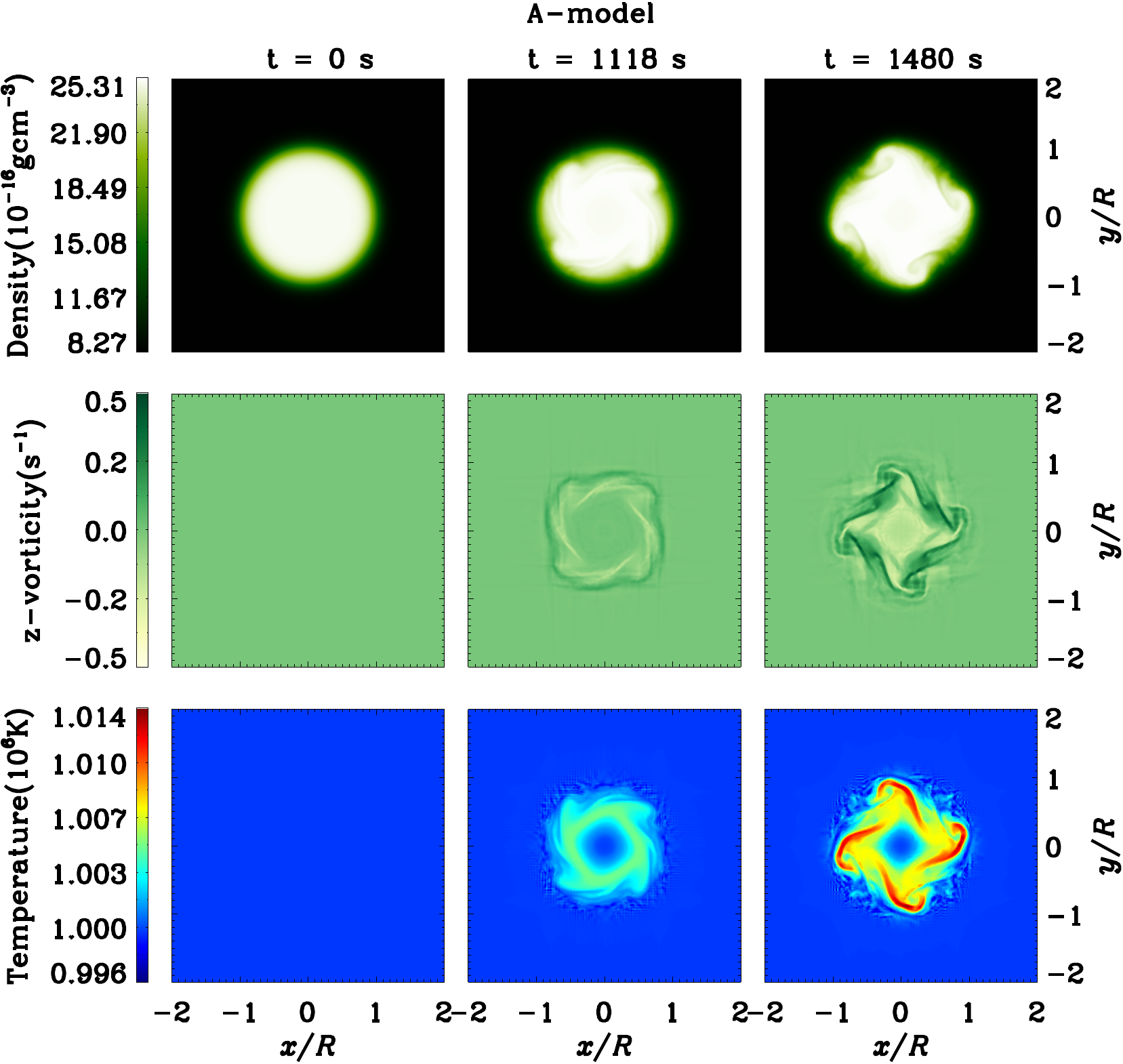}{0.33\textwidth}{(b)}
          \fig{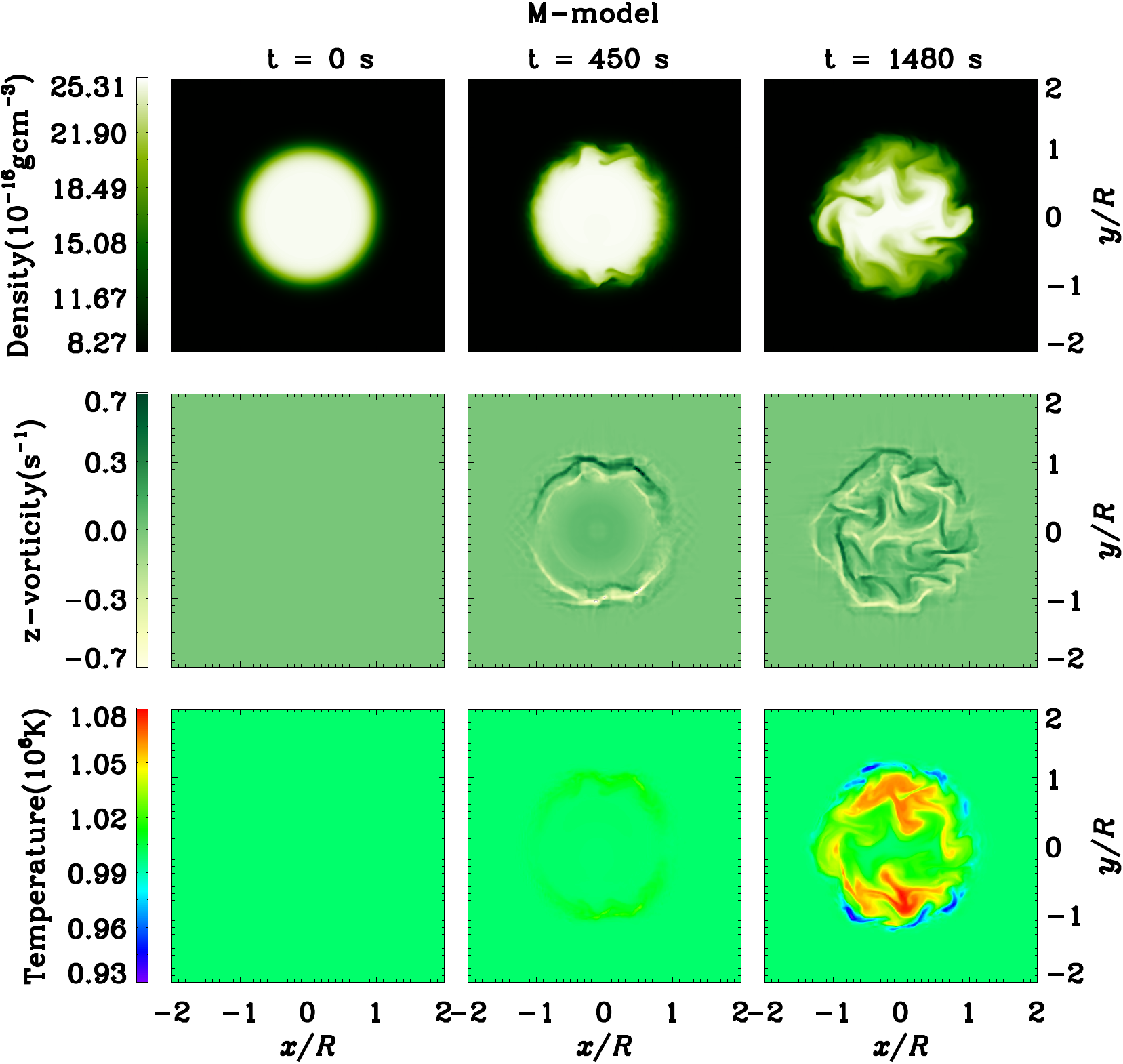}{0.33\textwidth}{(c)}
          }
          
\caption{Snapshots of density (upper row), $z$-vorticity (middle row) and temperature (lower row) evolutions of the cross-section at the loop apex 
for the K-model (a), the A-model (b) and the M-model (c). 	}			
\label{fig_rho_tem}
\end{figure*}

The simulation results show that the loops quickly form driven standing waves in the three models, 
namely standing kink (Alfv{\'e}n) waves in the K-model (A-model) and mixed (both standing kink and Alfv{\'e}n) modes in the M-model. 
As in previous studies,
the generation of KHI can also be seen in our K-model, 
as is shown in Figure \ref{fig_rho_tem}(a). 
The KHI develops near $|y|=R$, 
inducing the so-called TWIKH rolls \citep{2014ApJ...787L..22A,
2017ApJ...836..219A}.
Figure \ref{fig_rho_tem}(b) shows that axisymmetric vortices occur around the loop boundary in the A-model.
This means that the Alfv{\'e}n oscillations in a non-uniform layer can also induce the instability, 
which corresponds to the prediction of \citet{1983A&A...117..220H}.
We can observe four clear eddies around $|y|=R$ at $t=450{\rm s}$ in the K-model.
Actually there are still four small eddies beside the clear ones around the loop boundary,
which can be observed in the later instant ($t=1480{\rm s}$).
This means that the initial unstable mode in the K-model has a wavenumber of $m=8$.
In the A-model,
four eddies start to occur at $t=1118{\rm s}$,
indicating that the wavenumber of the initial unstable mode is $m=4$.

The results of the M-model are shown in Figure \ref{fig_rho_tem}(c).
It is almost the same snapshot as in the K-model at $t=450{\rm s}$,
indicating that the torsional motions inside the loop have little influence on the instability near the loop boundary initially.  
When the instability induced from the torsional waves develops,
the loop is deformed and the eddies extend from the boundary to almost the whole region in the M-model,
as is indicated by the $z$-vorticity in the middle panel of Figure \ref{fig_rho_tem}.
Note that the results here are attributed to not only the effect of mixed motions,
but also a higher energy input in the M-model than in the other two models. 

We also plot the temperature evolution of the apex in the bottom row in Figure \ref{fig_rho_tem}.
The temperature increases at the locations where small eddies develop for all three models.
Meanwhile, we can also observe a temperature decrease around the boundary edges.
The fluctuations of the temperature probably do not mean that the heating indeed happens at those small spatial-scale structures.
This property is explained as adiabatic heating (cooling) rather than real dissipation \citep{2015A&A...582A.117M,2017ApJ...836..219A, 2018ApJ...856...44A, 2017A&A...604A.130K}.
It should be noted that the temperature increase in the A-model is smaller than in the other two models.
This means that the Alfv{\'e}n modes do not produce so many small eddies to deform the loop,
therefore the density has a smaller change,
leading to a smaller temperature increase.

\begin{figure}
\centering
\includegraphics[width=0.5\columnwidth]{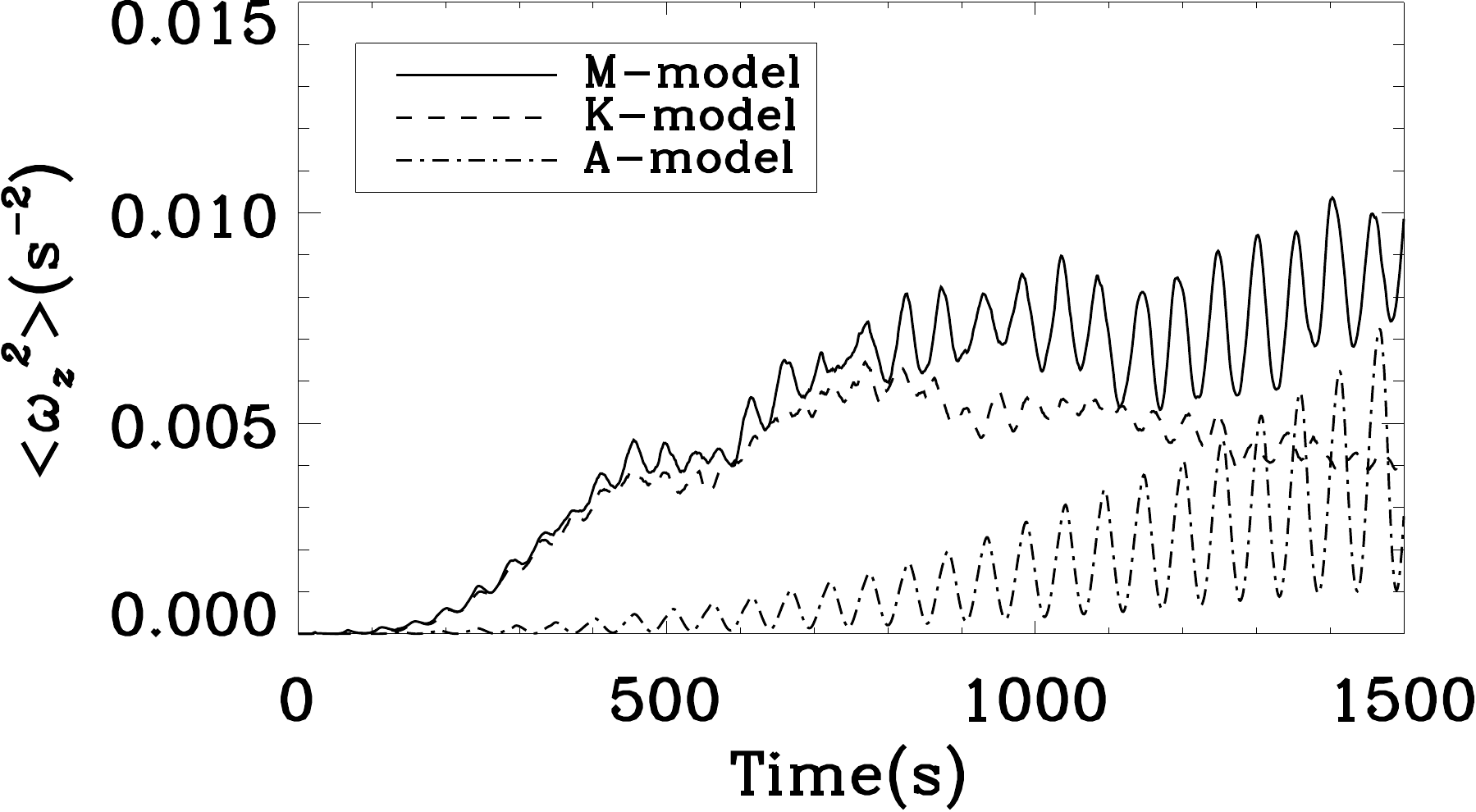}
\caption{Time evolution of the averaged square $z$-vorticity ($\omega^2_z$) for the M-model (solid line), 
K-model (dashed line) and A-model (dot-dashed line). 
The quantities are averaged over the region of $|x,y|\leq 2R$ at the loop apex. }	
\label{fig_vorticity}				
\end{figure}

To quantify the turbulent level in our models,
we examine the averaged square $z$-vorticity ($\omega^2_z$) at the loop apex,
which is shown in Figure \ref{fig_vorticity}.
The $\omega^2_z$ in the M-model is the largest,
indicating that the instability in this model is the strongest.
However, 
the amplitude increase of $\omega^2_z$ in the A-model does not mean that more eddies are generated in this model.
It is mainly due to the increasing torsional motions at the loop apex.

\begin{figure*}
\gridline{\fig{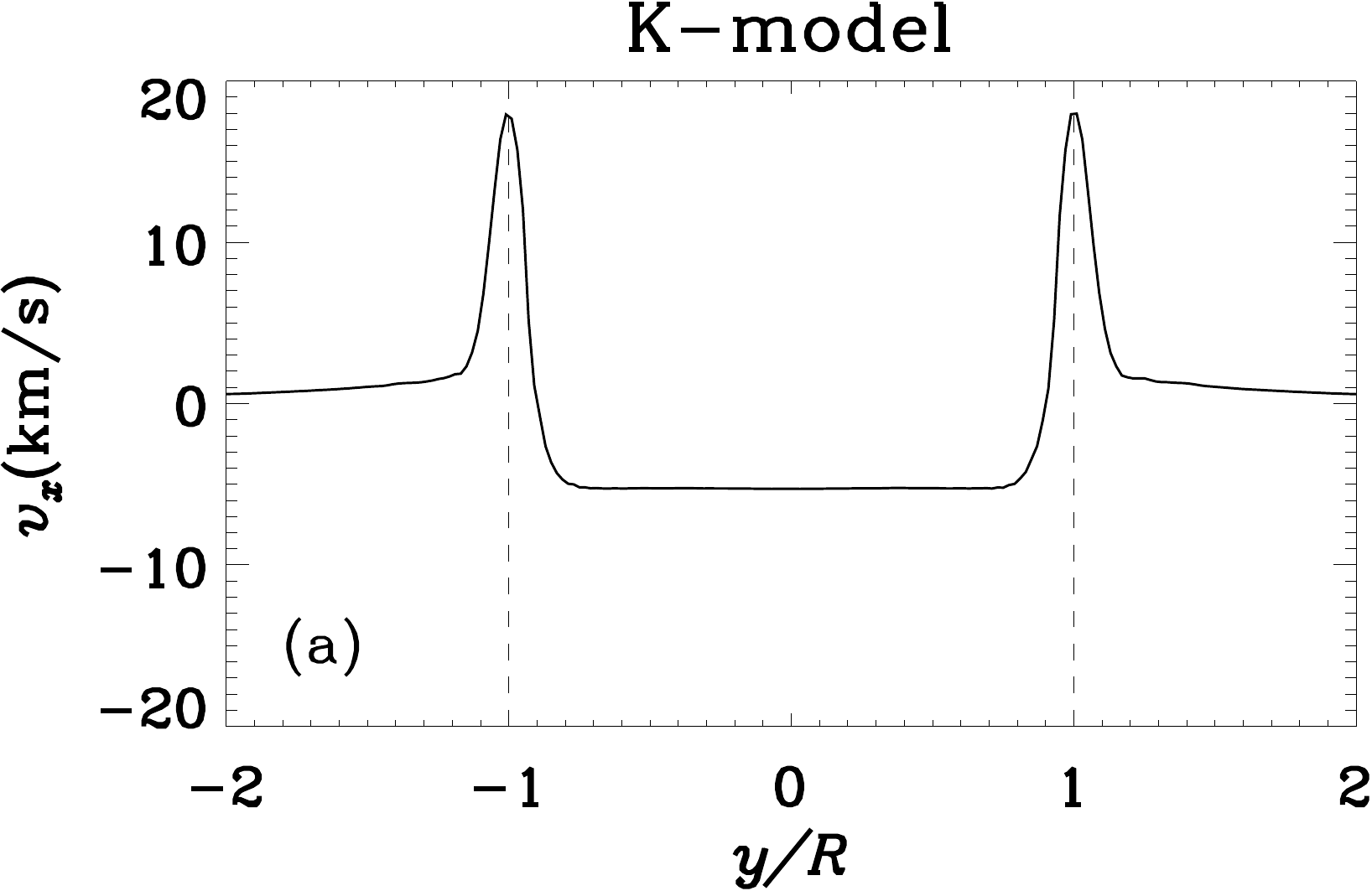}{0.35\textwidth}{}
          \fig{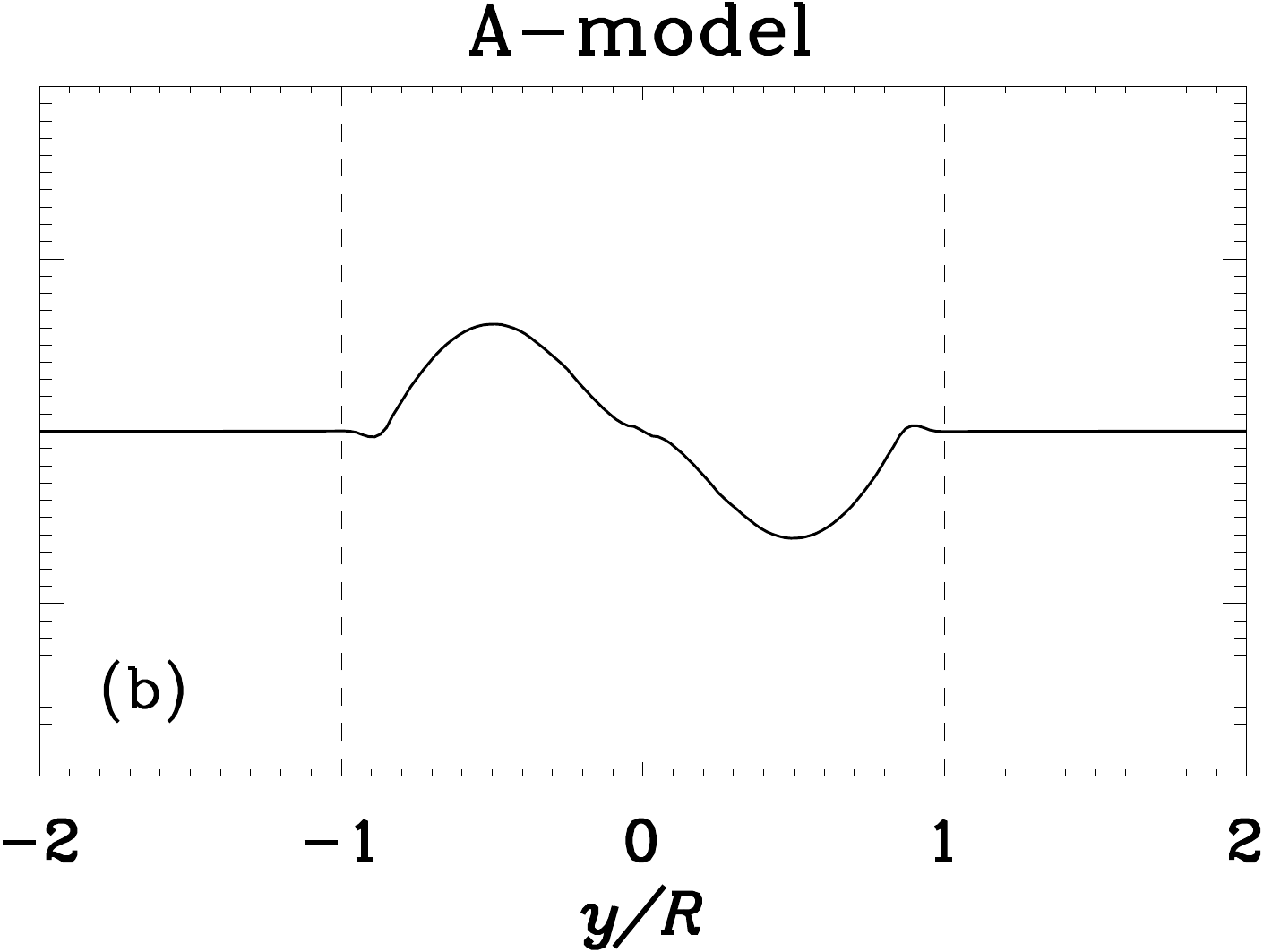}{0.3\textwidth}{}
          \fig{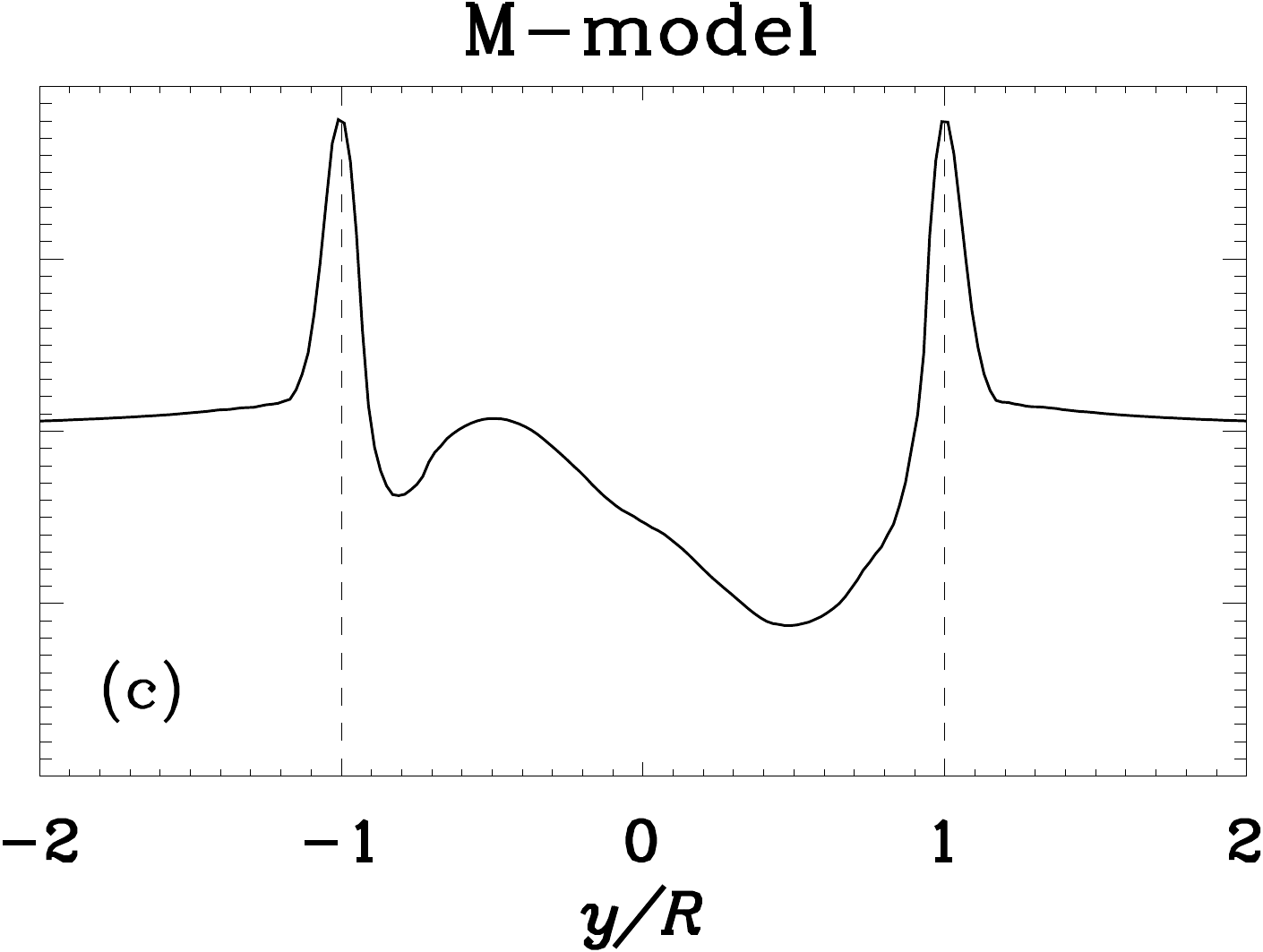}{0.3\textwidth}{}
          }
\caption{$v_x$ profile along the $y$-direction at $x=0$ and at the apex of the loops 
for the K-model (a), A-model (b) and M-model (c) at $t=255{\rm s}$. 
Vertical dashed lines represent the locations of loop boundaries.}	
\label{fig_resonant_layer}				
\end{figure*}

Alfv{\'e}n modes converted from kink oscillations through resonant absorption can be easily seen near the loop boundaries \citep{1988JGR....93.5423H, 1992SoPh..138..233G,2002A&A...394L..39G}.
Figure \ref{fig_resonant_layer} (a)(c) shows
the velocity spikes near $|y|=R$ in the K-model and the M-model at $t=255{\rm s}$.
The spikes are the Alfv{\'e}n modes converted from kink oscillations.
We do not find the Alfv{\'e}n modes at the same locations in the A-model in Figure \ref{fig_resonant_layer} (b),
because no kink oscillations appear in this model.
The crests near $y=-0.5R$ and the troughs near $y=0.5R$ in the A-model (Figure \ref{fig_resonant_layer} (b)) 
and the M-model (Figure \ref{fig_resonant_layer} (c)) are the Alfv{\'e}n oscillations coming from the drivers.
It should be noted that in the M-model, 
the Alfv{\'e}n oscillations inside the loops can mix with the Alfv{\'e}n modes in the nonuniform layer due to their different periods,
inducing the KHI.
So small structures can be seen near $y=-0.8R$ in Figure \ref{fig_resonant_layer} (c).

 The Alfv{\'e}n oscillations with different frequencies can have phase mixing after a number of periods \citep{1983A&A...117..220H}.
However,
the scales of phase mixing eddies will decrease over time,
since phase mixing will generate larger gradients and smaller scale structures.
According to \citet{1995JGR...10019441M} \citep[see also][]{2015ApJ...812..121K, 2017A&A...602A..75R}, 
the finest scale structures are governed by the phase mixing length
\begin{equation}
L_{\rm ph}=\frac{2L}{t(v_{\rm Ae}-v_{\rm Ai})/l} .
\label{eq_Lph}
\end{equation}
For a very late instant $t={\rm 1480s}$, 
the phase mixing length is $L_{\rm ph}={\rm 0.039Mm}$, 
which is already very close to our numerical resolution.
So we can not clearly observe such small structures any more.
Besides,
the onset of the instability can also make the identification of the phase mixing fine structures become ambiguous.

More eddies occur in the M-model,
indicating that the mixed torsional and transverse motions deform the loop efficiently.
Meanwhile, considering the small structures induced by phase mixing,
we find that the mixed modes are more efficient in generating such small spatial-scale structures. 
Therefore, 
they are also likely to dissipate energy into heating more efficiently.

\subsection{The saturation of oscillations}
\label{sec_sub_satu}

\begin{figure*}
\gridline{\fig{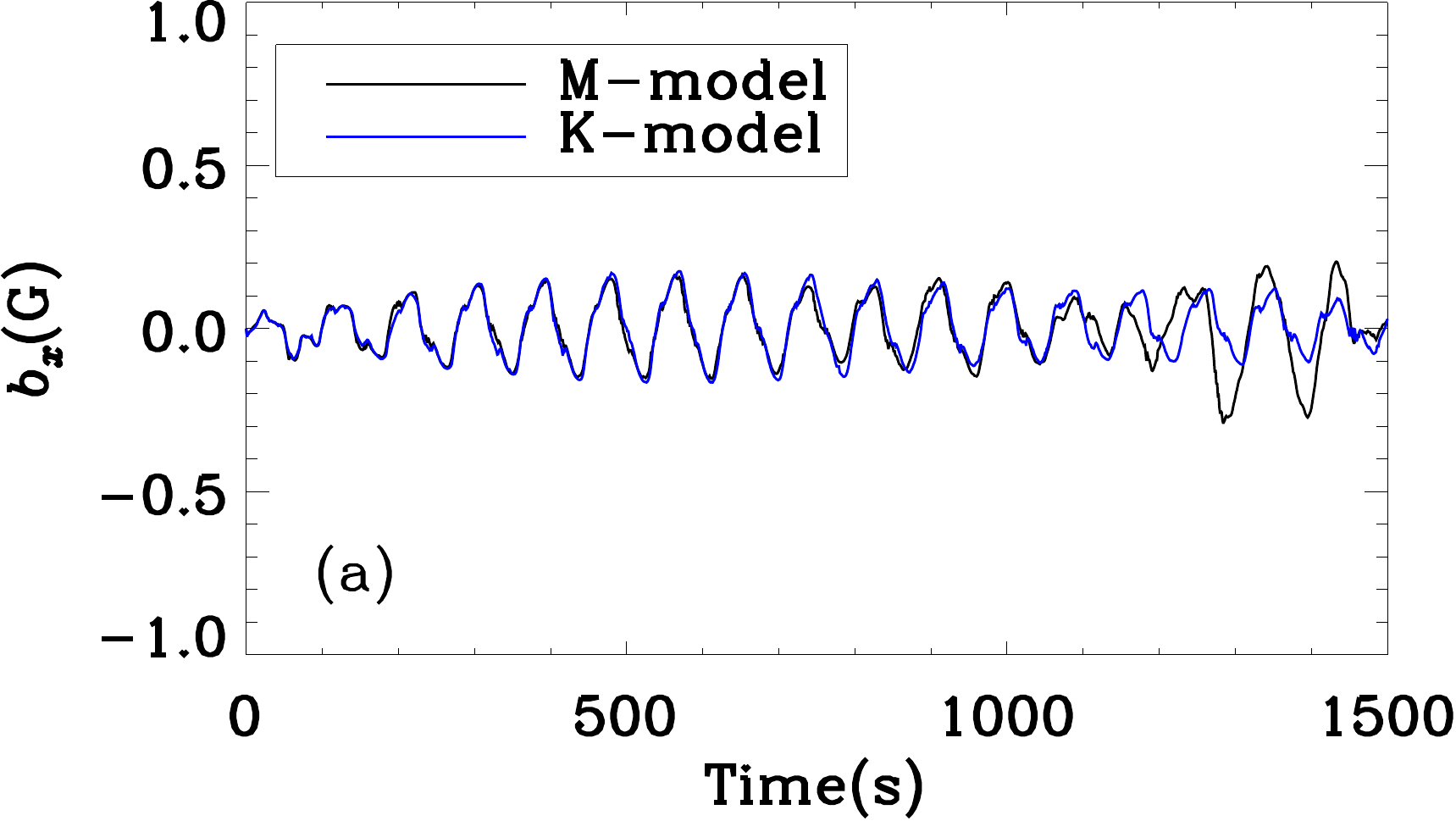}{0.48\textwidth}{}
          \fig{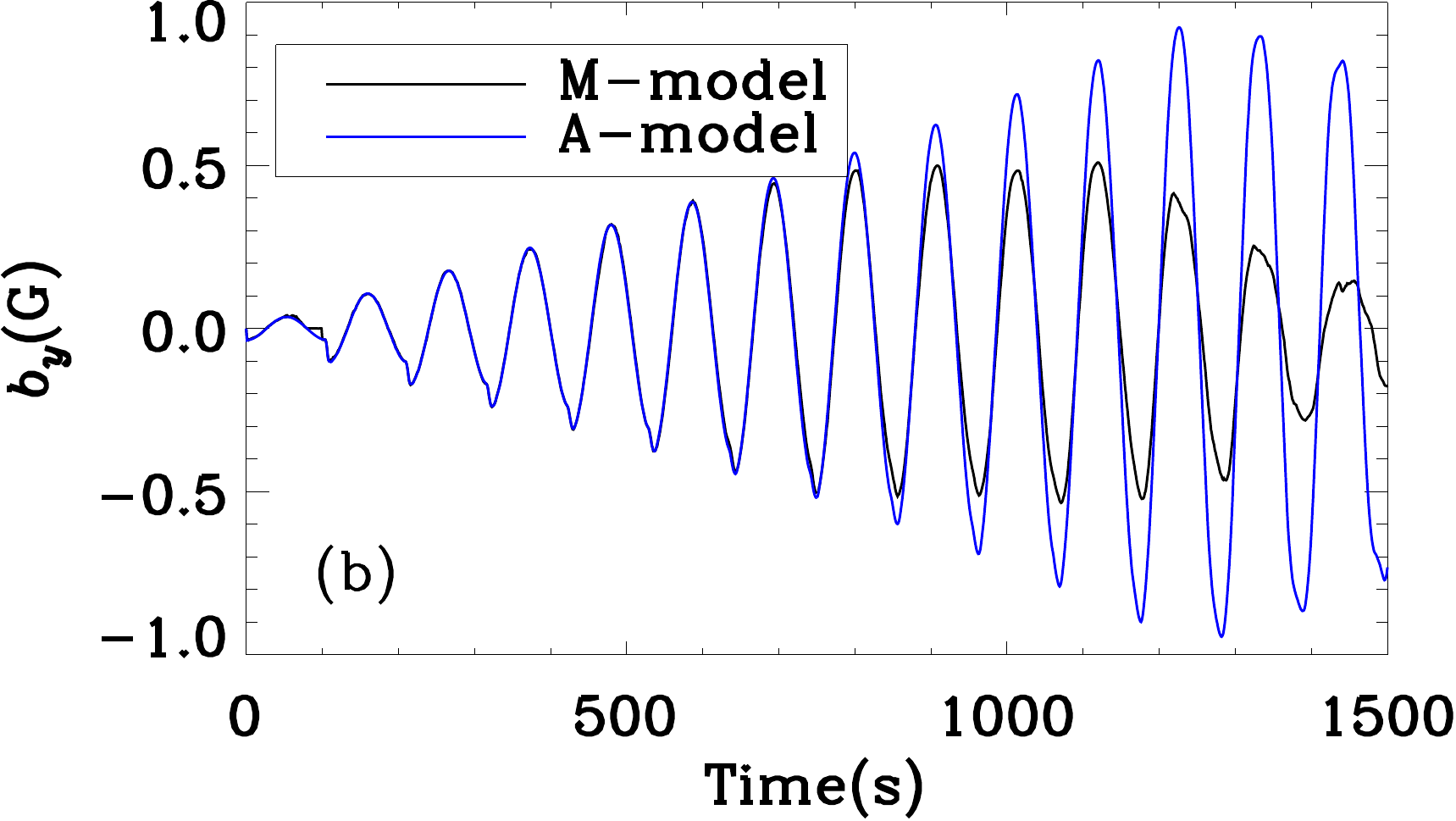}{0.48\textwidth}{}
          }
\caption{Transverse magnetic field perturbations at [$0.5R$,0,0] in the three models. 
The point is a fixed one which is not advected according to the drivers.
The left panel shows the $b_x$ evolution for the M-model(black line) and the K-model (blue line).
The right panel shows the $b_y$ evolution for the M-model (black line) and the A-model (blue line).}
\label{fig_bxy}				
\end{figure*}

Once we set up fundamental kink oscillations in loops, 
the direct approach is to check the displacements or velocities at the apex, 
which is the location of the antinode of transverse motions.
However,
because of the deformation of loops, 
the displacement of the apex can not reveal the true oscillation properties any more.
Although the deformations of our loops are not as strong as \citet{2018A&A...610L...9K} due to our smaller period and larger magnetic field,
to avoid the influence of the deformation,
we choose the perturbations of the transverse magnetic field at the footpoint to examine the oscillation properties.
For fundamental oscillations in loops,
the perturbations of the transverse magnetic field will have its maximum values at the footpoints.
Figure \ref{fig_bxy} shows the transverse perturbations of the magnetic field at the point [$0.5R$,0,0] in the three models. 
The specific point here is fixed at the bottom plane, 
so it is not advected following the drivers.
The maximum displacement of the central loop region in this plane is about ${\rm 27km}$, 
which is close to our resolution of $20 {\rm km}$.
So considering a fixed point does not significantly influence the results.

Figure \ref{fig_bxy} (a) shows the profiles of $b_x$ in the M-model and the K-model.
Since the Alfv{\'e}n motions do not have $x$-component inside the loops,
the $b_x$ here mainly represents the kink motions.
The amplitudes of $b_x$ in the two models are identical before $t=1100{\rm s}$,
showing that the kink oscillations are formed in both models
and they quickly achieve a same saturation after about 3 periods due to resonant absorption.
However, 
the Alfv{\'e}n modes need a longer time to saturate,
leading to larger saturation values,
as is shown in Figure \ref{fig_bxy} (b).
The amplitudes of $b_y$ in the M-model and the A-model are identical before $t=800{\rm s}$.
Then the saturation comes after that in the M-model,
while it saturates after about $t=1200{\rm s}$ in the A-model.
It should be noted that the amplitude of $b_x$ in the M-model increases after about $t=1200{\rm s}$,
whereas the amplitude of $b_y$ decreases after about $t=1200{\rm s}$.
This is because the point chosen here is very close to the edge of an eddy,
which makes the magnetic field vector component in the bottom plane deflect to the $x$-direction.

\section{ENERGETICS}
\label{sec_energy}

To understand the energy transfer in the systems, 
we study the time evolution of different kinds of energy.
In the following parts, 
we will analyse volume averaged values in the subregion of $|x| \leq 2R, |y| \leq 2R, 0\leq y \leq L$.
The input energy, 
namely the Poynting flux provided by the driver, is calculated by 
\begin{equation}
S(t)=-\frac{1}{V}\int^t_0\int_A \textbf{S}\cdotp d\textbf{A}dt',
\label{eq_poynting}
\end{equation}
following the definition in \citet{1999ApJ...526..478B}. 
Here \textbf{S} is the Poynting flux, 
\textbf{A} is the normal surface vector of the bottom plane
 and $V$ is the total volume of the subregion. 

\begin{figure*}
\gridline{\fig{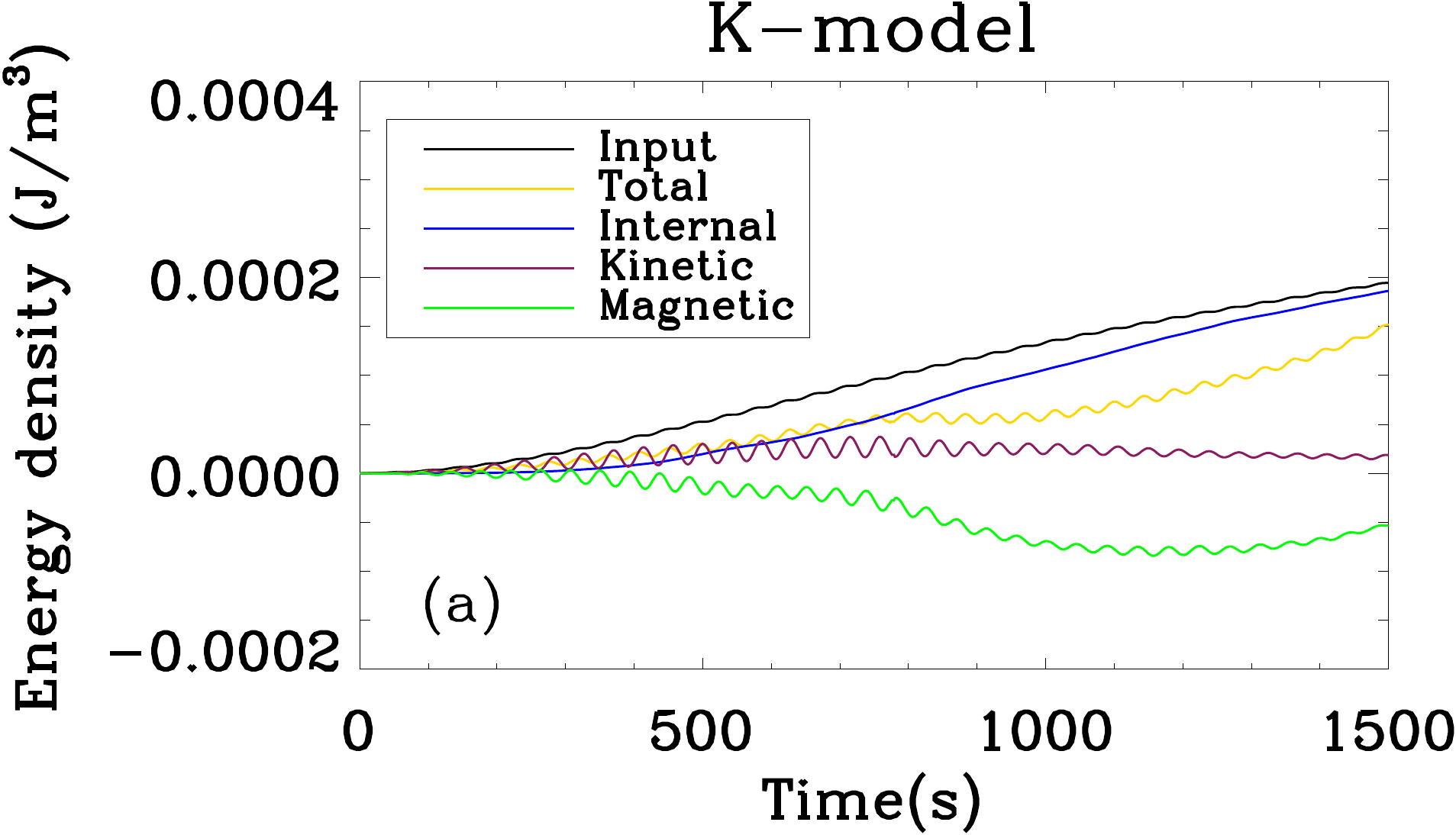}{0.38\textwidth}{}
          \fig{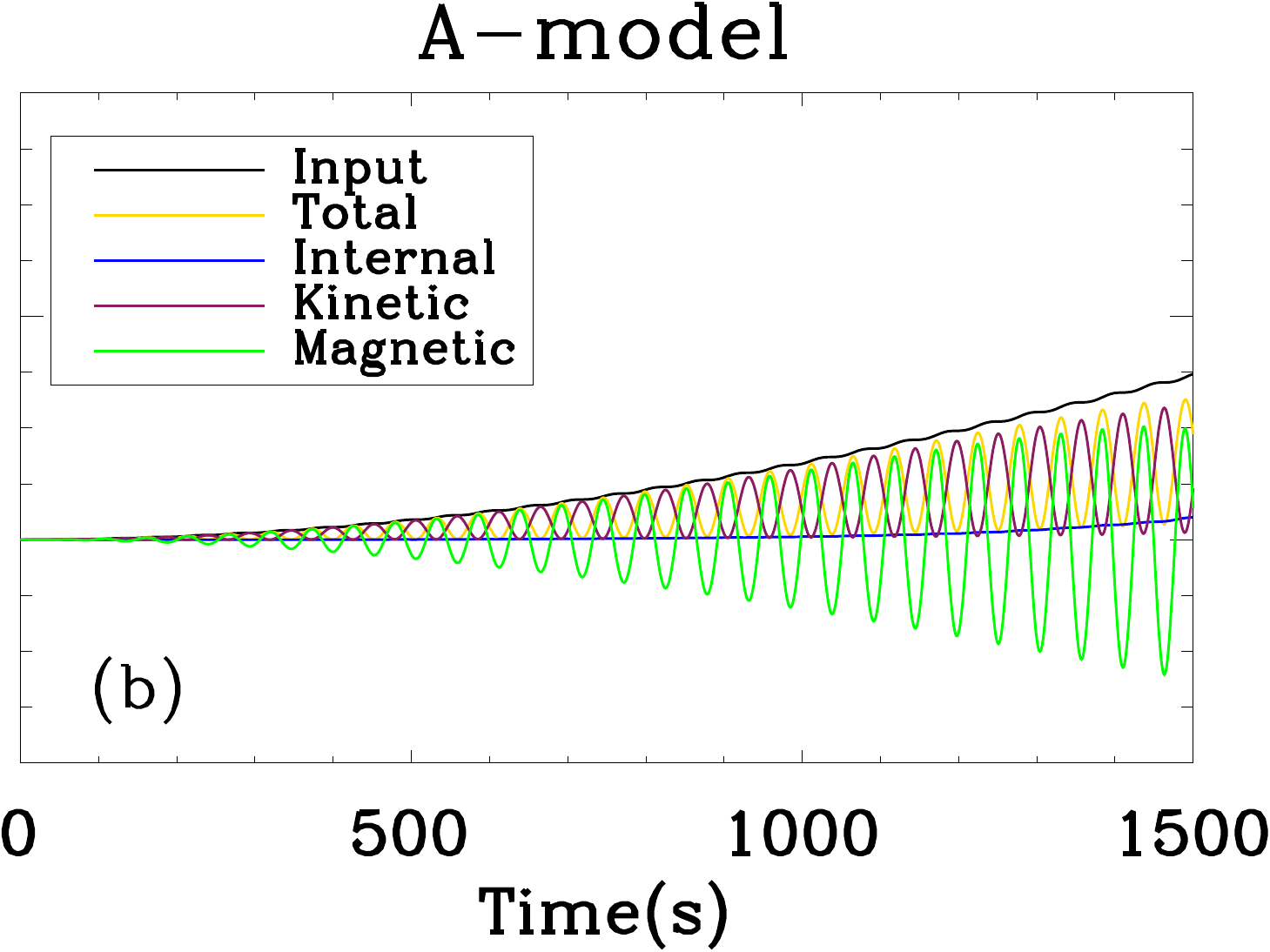}{0.29\textwidth}{}
          \fig{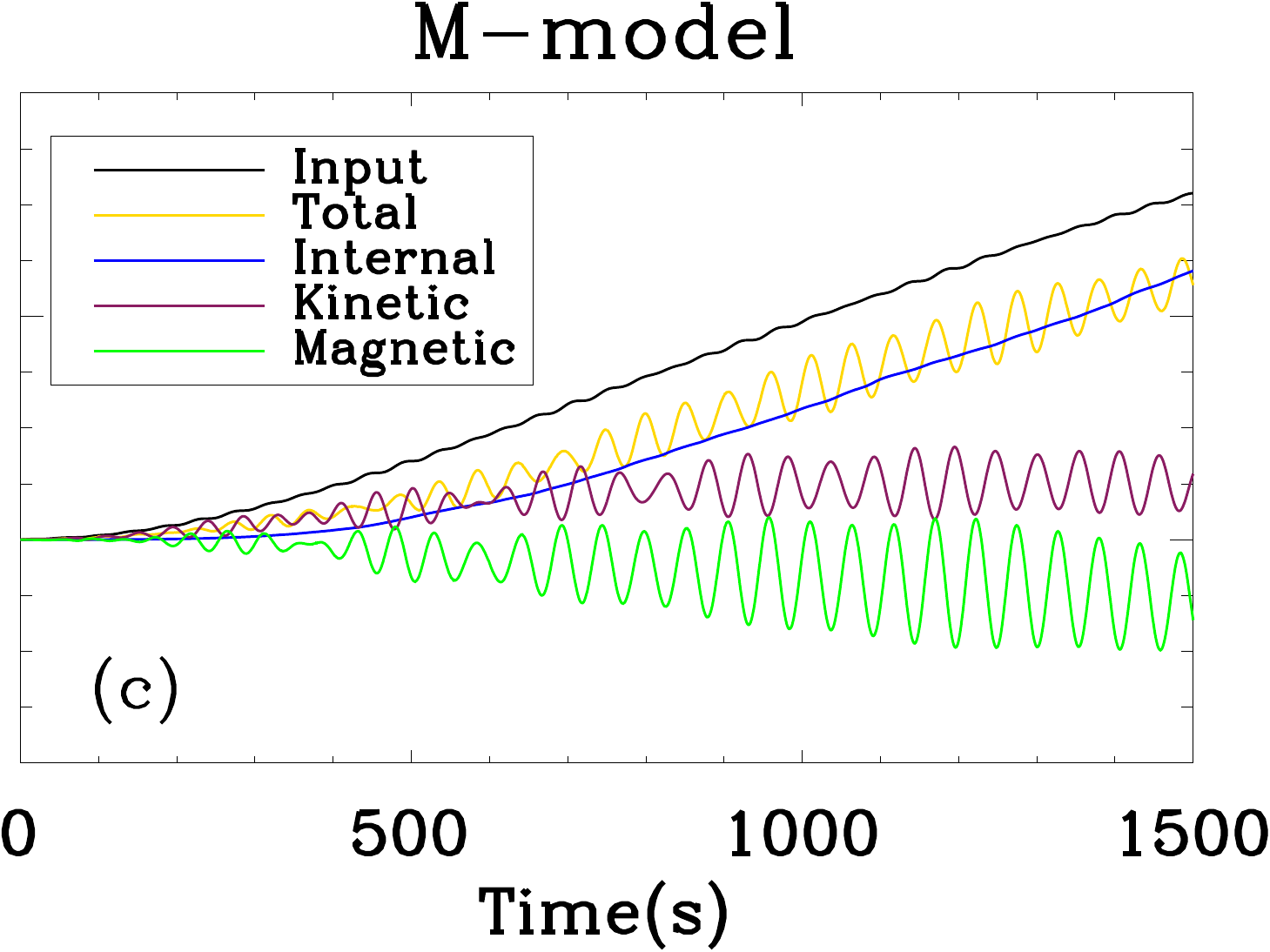}{0.29\textwidth}{}
          }
\caption{Volume averaged energy density variations relative to the initial state for the K-model (a), A-model (b) and M-model (c).
The energy densities are volume averaged over the region of $|x,y|\leq 2R, 0\leq y\leq L$.
Different colours represent different kinds of energy density variations.
Note that the total energy density means the sum of internal, kinetic and magnetic energy density.}	
\label{fig_energy}				
\end{figure*}

Since all the variations are averaged in the same sub-volume, 
we will discuss energy instead of energy density in the following.
In Figure \ref{fig_energy}, 
the input energy for each model is approximately divided between the internal energy and the kinetic energy.
 In the K-model, 
a quick saturation in the kinetic energy is achieved,
with a slight decrease after 10 periods.
This is because the collective transverse oscillation transfers into the local turbulent motions near the loop boundary,
then the TWIKH rolls break up into smaller and smaller structures.
Meanwhile,
considering the extension of non-uniform layer \citep{2017A&A...604A.130K},
 these fine structures spread over a larger region,
 causing the decrease of the averaged velocity.
 Similar reduction in the vorticity of the K-model can also be observed in Figure \ref{fig_vorticity}.
Because of the decrease of the magnetic field perturbation in the bottom plane,
the input energy experiences a slower increase in the later periods in the K-model.
In the A-model and the M-model, 
both the kinetic energy and the magnetic energy have larger relative amplitudes,
indicating later saturations.
Note that in the M-model,
 beatings can be seen in the amplitudes of the kinetic energy and the magnetic energy due to the periods mismatch between the transverse and torsional waves.

The drop in the magnetic energy, 
namely the difference between the input energy and the total energy,
is caused mainly by the Poynting flux through our open lateral boundaries. 
Meanwhile,
the other part can be attributed to the energy transfer from the magnetic energy to the internal energy due to the effect of numerical resistivity.
This is similar to the results of \citet{2017A&A...604A.130K}.
We also notice a small rise near the end of our simulation for magnetic energy in the K-model, 
which is also mentioned by \citet{2017A&A...604A.130K}.
This is due to the continuous energy input of the driver.

The input energy is almost at the same level in the K-model and the A-model,
however,
the internal energy increase in the A-model is much smaller than in the K-model.
As is mentioned in Section \ref{sec_sub_eddies},
the pure torsional motions produce less eddies in the A-model.
Therefore,
the wave energy is less dissipated,
showing a very weak heating here.

In Figure \ref{fig_energy} (c), 
the increased input energy in the M-model becomes approximately proportional to time after about $t={\rm 500s}$.
We estimate the energy flux $E=\Delta S(t)V/\Delta tA\sim36.5 {\rm W m^{-2}}$, choosing a period from ${\rm 700s}$ to ${\rm 1200s}$.
This energy flux seems to get close to balance the radiative energy losses of quiet corona, $\sim100 {\rm W m^{-2}}$ \citep{1977ARA&A..15..363W, 2007Sci...317.1192T}.
Furthermore, 
it should be noted that the input energy flux would increase for a larger input velocity.
If we consider a larger amplitude driver, 
for example $4{\rm km~s^{-1}}$, 
the input kinetic energy would become four times larger, 
which would be enough to heat at least the quiet corona. 
Such a larger amplitude could be representative of driving velocities in e.g. the chromosphere.

\begin{figure}
\centering
\includegraphics{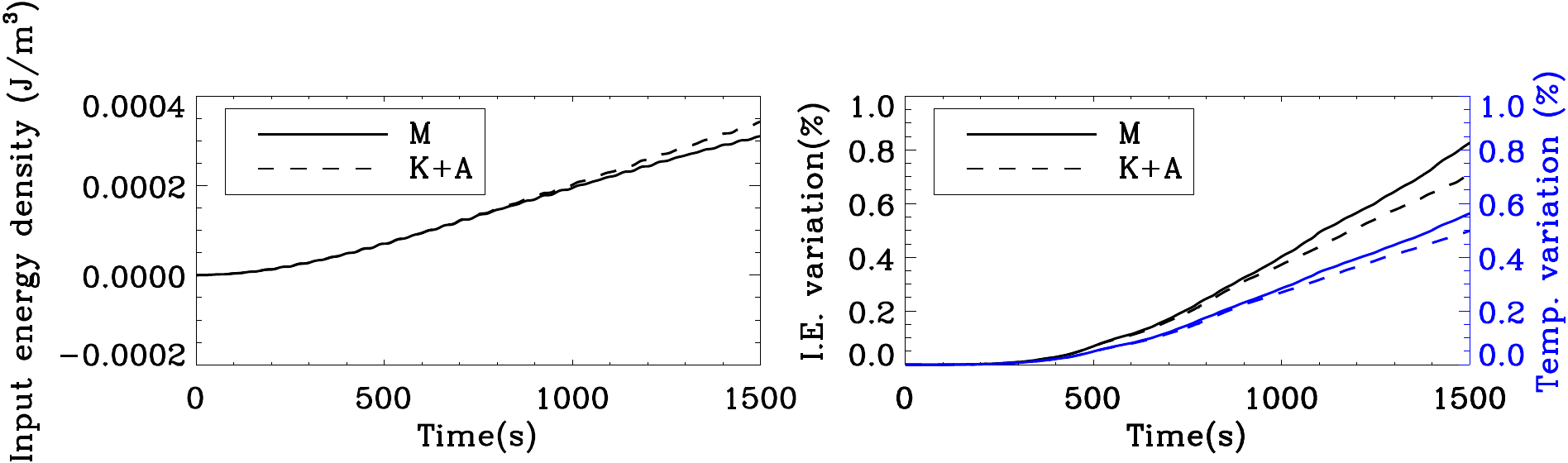}
\caption{Left: Volume averaged input energy density variations.
Right: Percentages of volume averaged internal energy (black) and temperature (blue) variations. 
Solid lines represent the M-model, dashed lines represent the sum of the K-model and the A-model.
The quantities are volume averaged over the region of $|x,y|\leq 2R, 0\leq y\leq L$. }	
\label{fig_aver_energy}				
\end{figure}

To clearly compare the variations of internal energy and temperature in all three models, 
we examine the percentages of volume averaged values.
Before that,
we compare the input energy in the M-model and the sum of the other two models,
as is shown in the left panel of Figure \ref{fig_aver_energy}.
They are identical before $t=1000{\rm s}$,
then the input energy in the M-model gets smaller than the sum of the other two models.
This is due to the decrease of the magnetic field perturbations near the footpoint in the M-model.
The right panel of Figure \ref{fig_aver_energy} shows that the relative variations of internal energy and temperature monotonically increase over time.
For the M-model, 
the relative variation of the internal energy increases to $0.83\%$ at the end of the simulation ($t=1500{\rm s}$).
Meanwhile, it increases to $0.71\%$ for the sum of the other two models.
Similarly, the relative variation of the temperature increases to $0.56\%$ at the end of the simulation for the M-model,
$0.49\%$ for the sum of the other two models. 
Although the input energy is even smaller in the later periods of the simulation,
both the internal energy and the temperature still get larger increases in the M-model.
This means that the mixed modes in the M-model can indeed have enhanced heating due to a more efficient dissipation than the other two models combined.
As such, 
the KHI rolls act as a catalyst to more efficiently dissipate the energy in other wave modes.

\section{OBSERVABLE PROPERTIES}
\label{sec_fomo}

To obtain observable signals and compare to real observations, 
we forward modelled the numerical simulations using the FoMo code \citep{2016FrASS...3....4V}. 
The Fe  \uppercase\expandafter{\romannumeral 9} 171 ${\rm \AA}$ emission line is chosen since it is sensitive to the temperature of the models here.

\begin{figure*}
\centering
 (a) K-model

\gridline{\fig{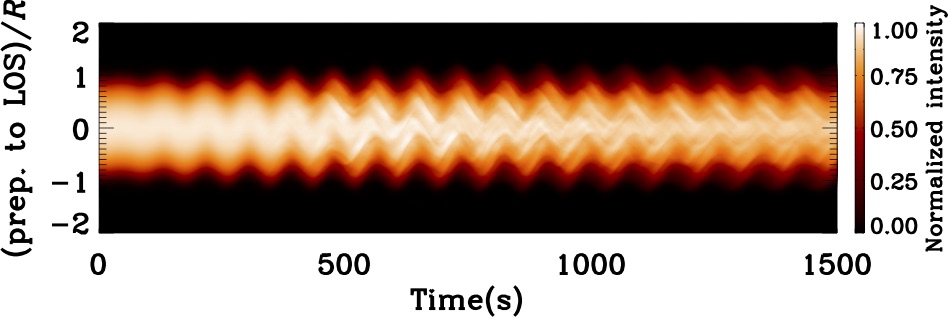}{0.32\textwidth}{}
          \fig{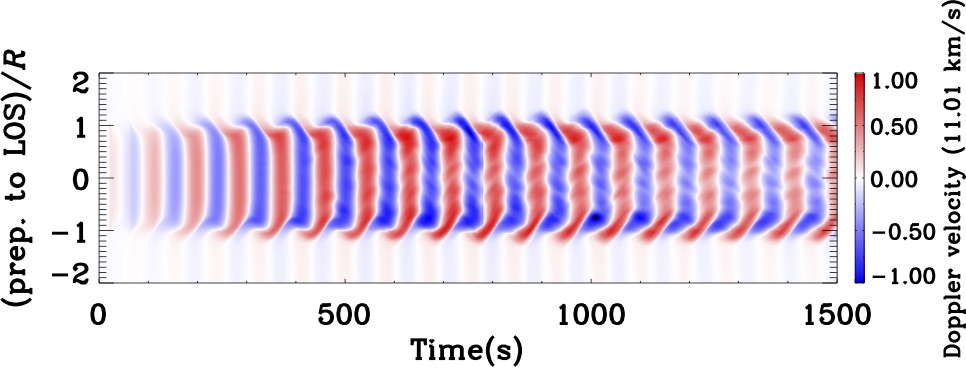}{0.32\textwidth}{}
          \fig{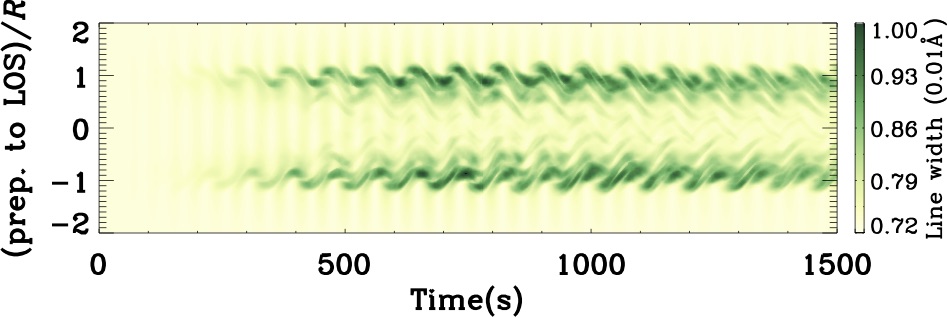}{0.32\textwidth}{}
          }
\gridline{\fig{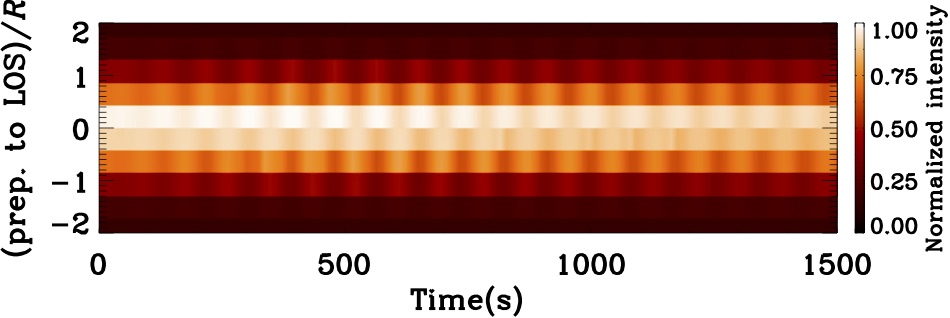}{0.32\textwidth}{}
          \fig{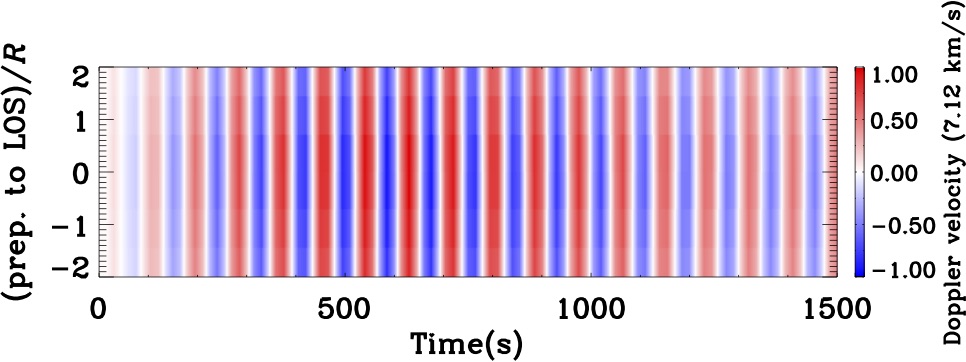}{0.32\textwidth}{}
          \fig{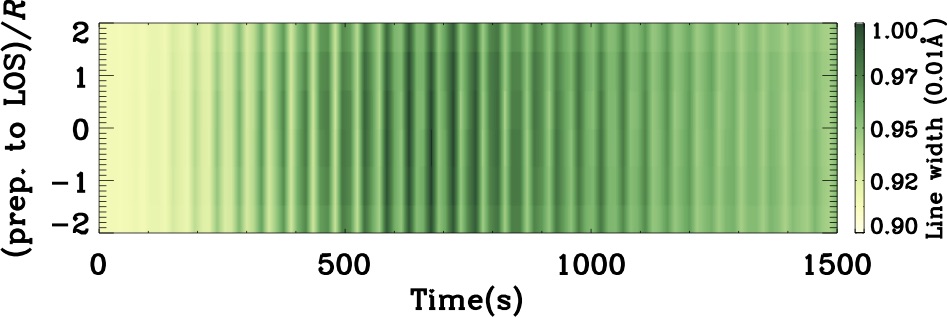}{0.32\textwidth}{}
          }
         
 \centering
 (b) A-model         
\gridline{\fig{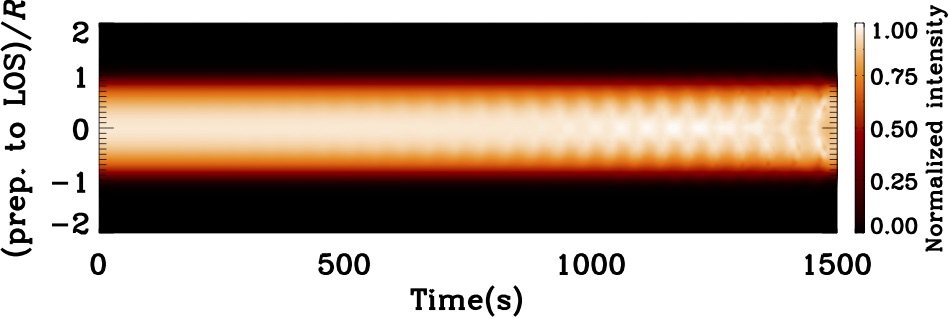}{0.32\textwidth}{}
          \fig{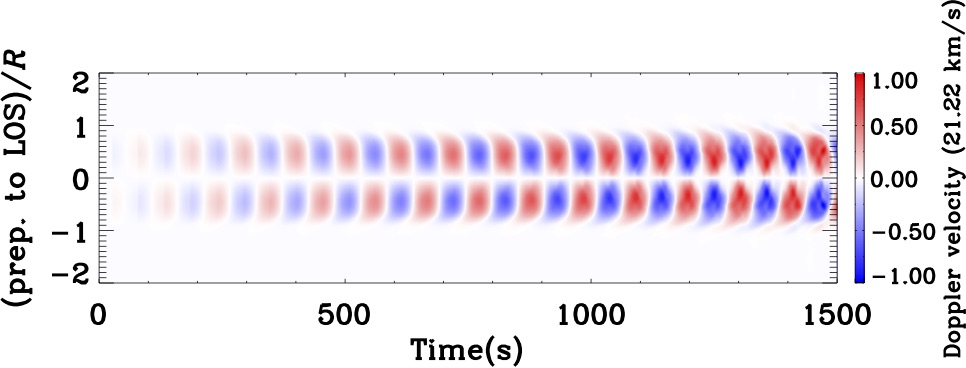}{0.32\textwidth}{}
          \fig{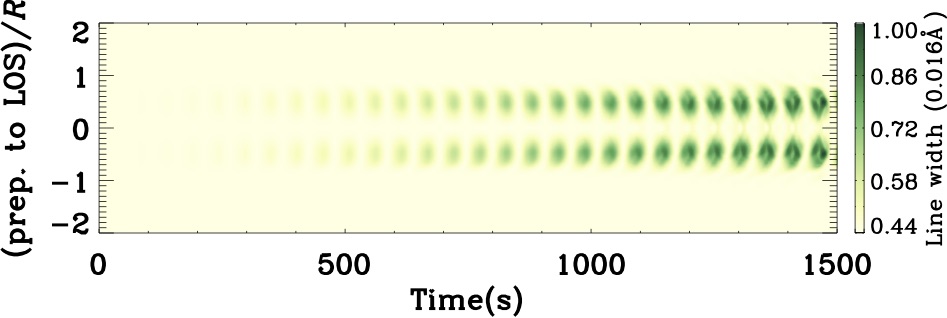}{0.32\textwidth}{}
          }
\gridline{\fig{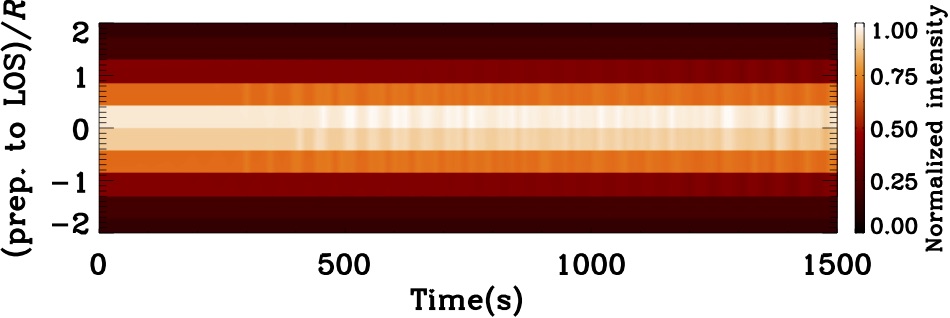}{0.32\textwidth}{}
          \fig{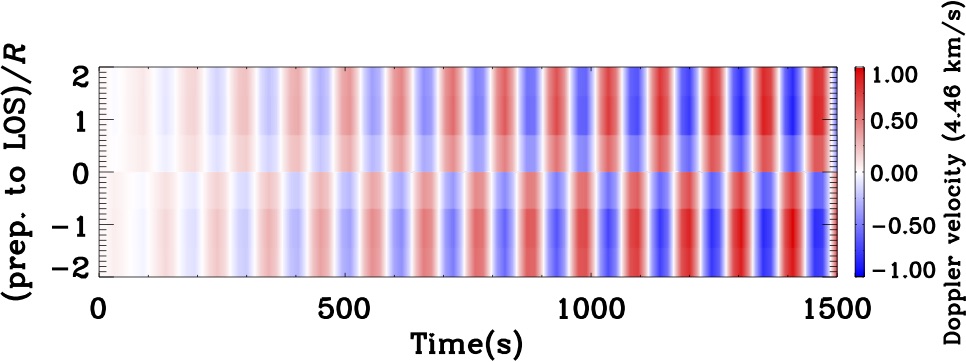}{0.32\textwidth}{}
          \fig{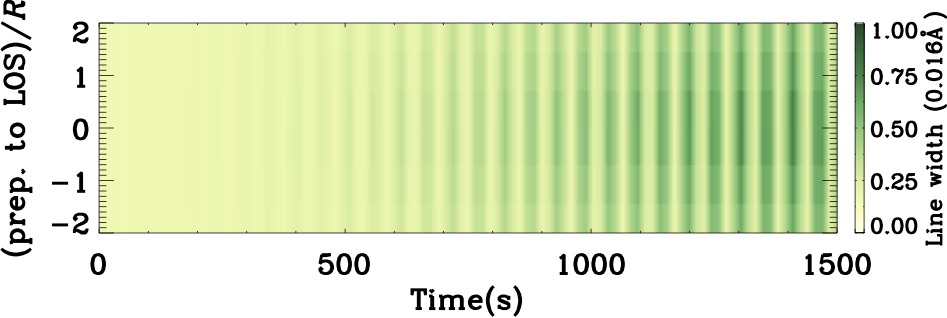}{0.32\textwidth}{}
          }
 \centering
 (c) M-model         
\gridline{\fig{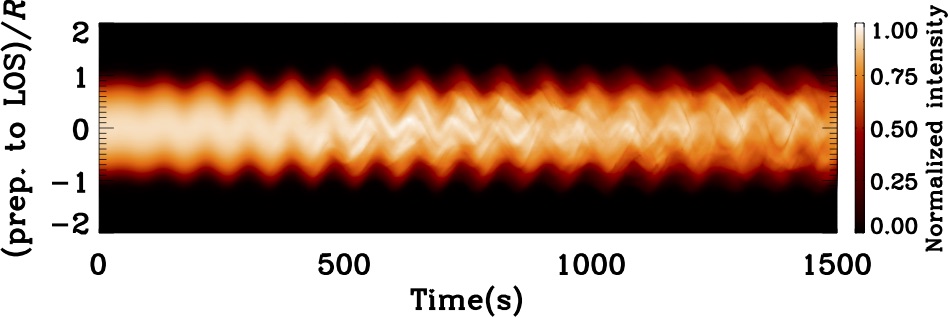}{0.32\textwidth}{}
          \fig{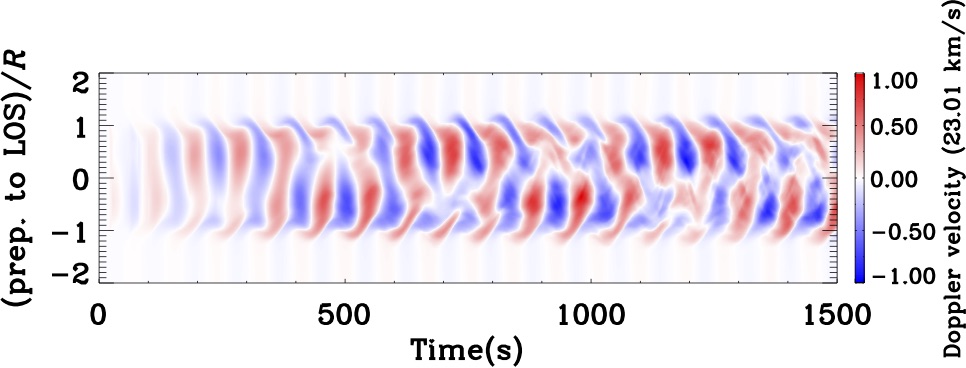}{0.32\textwidth}{}
          \fig{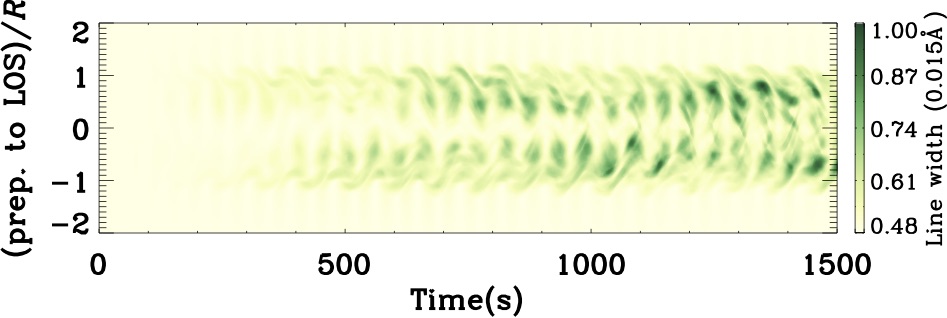}{0.32\textwidth}{}
          }
\gridline{\fig{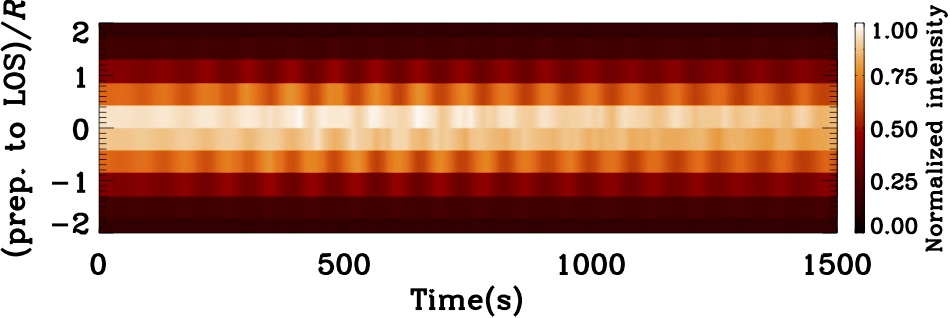}{0.32\textwidth}{}
          \fig{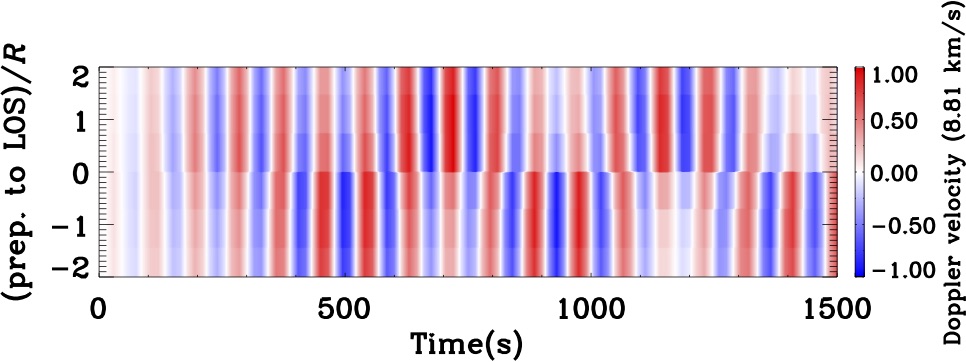}{0.32\textwidth}{}
          \fig{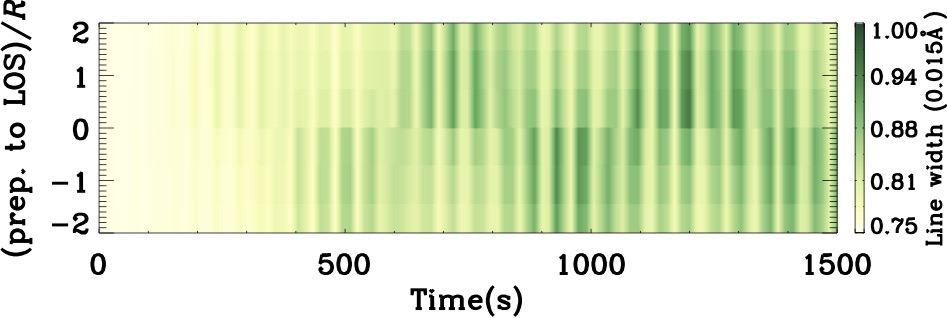}{0.32\textwidth}{}
          }

\caption{Forward modelling results for the three models in the Fe  \uppercase\expandafter{\romannumeral 9} 171 ${\rm \AA}$ line at the apex with a LOS angle of $45^\circ$.
The left panel of each model: Time-distance maps of the normalized intensities. 
The upper one is obtained with the full numerical resolution and the lower one with a degraded resolution comparable to SDO/AIA. 
The middle panel of each model: Time-distance maps of the normalized Doppler velocity. 
The upper one is obtained with the full numerical resolution and the lower one with a degraded resolution comparable to Hinode/EIS.
The right panel of each model: Similar to the Doppler velocity maps, but for the spectral line width. }	
\label{fig_fomo}				
\end{figure*}

Figure \ref{fig_fomo}(a) shows the results for the K-model, 
where the left column shows the time-distance diagram of the normalized intensity in the Fe \uppercase\expandafter{\romannumeral 9} 171 ${\rm \AA}$ line 
at the loop apex with a LOS angle of $45^\circ$ (LOS angle of $0^\circ$ along the $y$-direction) within the plane perpendicular to the loop axis. 
The upper image is obtained with the full numerical resolution and
the fine strand-like structures can be clearly seen due to the instability after about 450s,
which can also be seen in impulsively excited loops \citep{2014ApJ...787L..22A, 2016ApJ...830L..22A,2017ApJ...836..219A}.
Similar to the results of \citet{2016ApJ...830L..22A},
the periodic increase of intensity after about $500{\rm s}$ around the boundary is caused by the TWIKH rolls.
The intensity increase is more apparent here due to the continuous energy increase in our model. 
The fine structures dim after about $1000{\rm s}$,
as the eddies break into much smaller ones,
so the smaller structures can no longer be seen. 
To compare to the observations of decayless oscillations in coronal loops,
we degrade the original spatial resolution to the one of a given imaging instrument, 
namely SDO/AIA here.
The result is shown on the left bottom of Figure \ref{fig_fomo}(a).
It is very similar to the observations reported by \citet{2013A&A...560A.107A,2015A&A...583A.136A},
meaning that our simulation agrees with a decayless transverse oscillation.
The same decayless oscillations can also be seen in the models of \citet{2016ApJ...830L..22A} with a coarse instrument resolution.
  The middle column of Figure  \ref{fig_fomo}(a) shows the Doppler velocities in the same emission line and a LOS angle of $45^\circ$.
Staggered blue and red shifts appear, 
showing a series of ``bow-like'' shapes.
It should be noted that they are similar to the results of \citet{2017ApJ...836..219A},
the moving of crests opposite to the loop core and their troughs move in the same phase as the loop core.
Smaller structures can now be seen after about 1000s, 
which agrees with the above statements.
To compare with a real instrument,
we also degrade the original numerical spatial resolution to $3''$.
We use a spectral resolution of $36 {\rm km~s^{-1}}$ and a temporal resolution of $ {\rm 15s}$ to mimic Hinode/EIS.
The bottom row of the middle panel of Figure \ref{fig_fomo}(a) shows the ``bow-like'' shapes can not be detected any more, 
due to the limited resolution.
The shapes become staggered red and blue stripes.
The right column of Figure \ref{fig_fomo}(a) is similar to the middle panel but for the spectral line width. 
We can not see the obvious line broadening before about $t=200{\rm s}$.
This is because the initially formed oscillations in our loop have relatively small amplitudes,
causing indistinguishable broadening.
Then the small structures generate rapidly around the loop boundary,
showing a significant broadening.
Similarly, 
the degraded results mimicking Hinode/EIS are shown in 
the bottom row of the right panel of Figure \ref{fig_fomo}(a).
The fine structures can not be seen, only stripes are detectable.

Figure \ref{fig_fomo}(b) shows the forward model of the A-model. 
The original and degraded resolution results of the imaging models can be seen in the left panel of Figure \ref{fig_fomo}(b).
No transverse oscillations appear, 
largely because the azimuthal incompressible Alfv{\'e}n modes do not disturb density.
But intensity fluctuations can be seen in the original resolution results after $500{\rm s}$,
due to the KH instability induced through phase mixing.
The fluctuations are not visible in the degraded resolution.
Therefore,
the imaging instruments can not observe torsional Alfv{\'e}n modes at their current resolution.
The middle panel of Figure \ref{fig_fomo}(b) shows the Doppler velocity maps.
The original resolution results (upper row) present staggered spot regions,
showing that axisymmetric Alfv{\'e}n oscillations are set up in the loop.
Similar to the K-model case, 
staggered red and blue stripes appear in the degraded models. 
We can also find signatures of Alfv{\'e}n oscillations in the right panel of Figure \ref{fig_fomo}(b),
where the line width broadens inside the loop due to the torsional motions.
Only stripes are detectable in the bottom row of the right panel of Figure \ref{fig_fomo}(b) because of the coarse resolution.

Figure \ref{fig_fomo}(c) shows the forward modelled results of the M-model.
Fine structures can be seen in the intensity diagram as for the K-model.
After about ${\rm 800s}$, 
the structures seem more disordered, 
owing to the torsional motions of Alfv{\'e}n modes.
The same degradation procedure is done to mimic the observations of SDO/AIA.
Due to the limited resolution, 
neither the Alfv{\'e}n properties nor fine structures can be observed.
This diagram is similar to the K-model case and
they are both very similar to the real observations,
meaning that both models could provide explanations for the decayless oscillations.
The Doppler velocity map here also presents the blue and red shifts, 
but showing ``tadpole-like'' shapes.
The torsional Alfv{\'e}n waves break the ``bows'' into smaller ``tadpole'' pairs.
Due to the rotating and transverse motions,
superpositions happen at the ``heads'' and cancellations happen at the regions with no ``tadpole''.
In the bottom row of the middle panel of Figure \ref{fig_fomo}(c), 
with the resolution of Hinode/EIS,
the ``tadpole-like'' shapes can not be detected either,
red and blue stripes are generated instead.
The right column of Figure \ref{fig_fomo}(c) shows the line width maps.
As mentioned above,
the mixed wave modes can induce more turbulent structures.
Therefore, 
the line broadening can be observed in almost the whole loop region
and disordered broadening shapes can be seen.
Similar to the Doppler shift properties,
the fine structures in line width can not be observed in the lower row of the right panel of Figure \ref{fig_fomo}(c).
Considering the frequency mismatch between the kink modes and Alfv{\'e}n modes in the M-model,
we would expect a beating behaviour between these two wave modes.
As is shown in Figure \ref{fig_fomo}(c),
the increases in Doppler velocity and line width show beatings,
which can also be observed in the modulation of the kinetic and magnetic energy amplitudes in Figure \ref{fig_energy} (c).

\begin{figure}
\centering
\includegraphics[width=0.5\columnwidth]{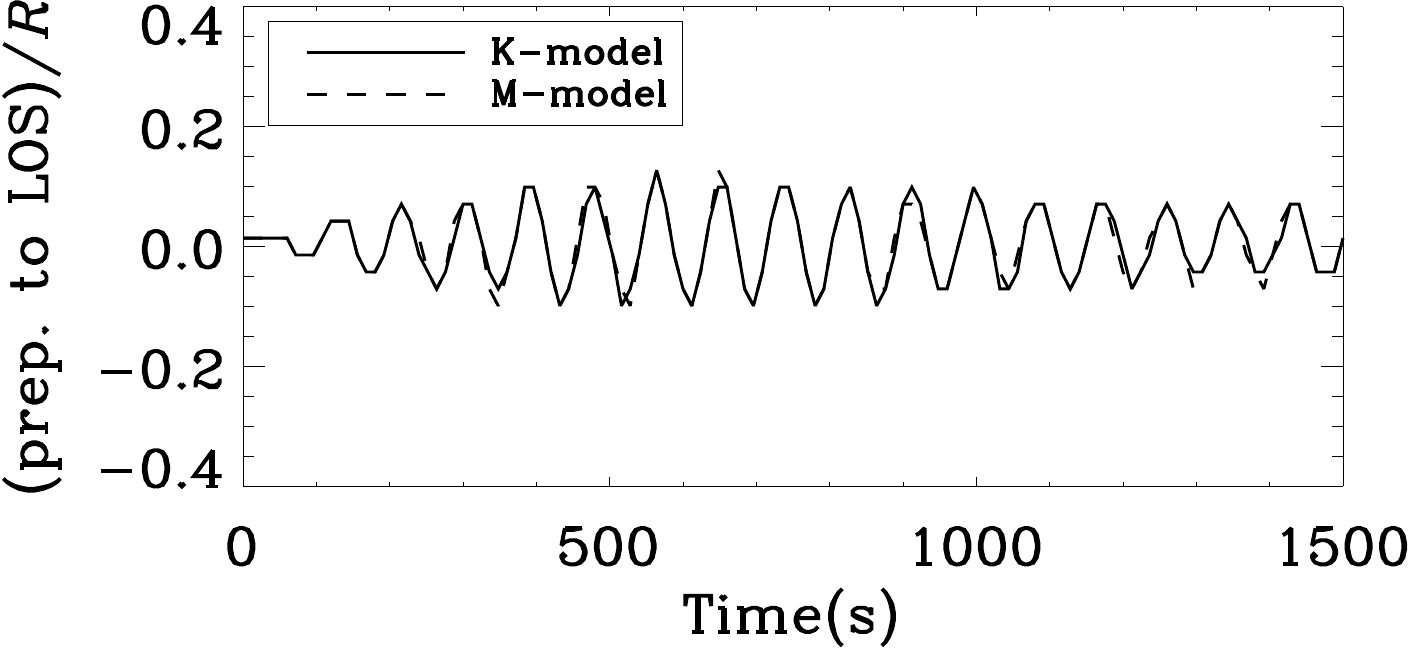}
\caption{Oscillation profiles of degraded resolution intensities,
obtained by calculating the maximum values of Gaussian fits of the results in the left bottom rows of Figure \ref{fig_fomo}(a) and Figure \ref{fig_fomo}(c).
Solid and dashed lines represent the results of the K-model and the M-model, respectively.}

\label{fig_fitting}				
\end{figure}

 We plot the oscillation profiles of the degraded resolution intensities, 
 as is shown in Figure \ref{fig_fitting}.
 The intensity profiles are the maximum values of Gaussian fits of the results in the left bottom of Figure \ref{fig_fomo}(a) and Figure \ref{fig_fomo}(c).
 The profiles of these two models do not have significant difference,
 indicating only kink period signals can be observed by SDO/AIA.
 The amplitude here is about 0.1 Mm, 
 which agrees with the observed values in \citet{2013A&A...560A.107A}. 
 
We note that the staggered pattern of Doppler velocity in the A-model sets a clear difference with the case in the K-model.
Actually,
this has not been detected yet with EIS.
It indicates that the amplitudes of torsional Alfv{\'e}n waves assumed inside the loop are probably larger than the real ones.
Besides, 
the more localised distribution of the torsional Alfv{\'e}n modes would also influence the Doppler velocity in the coarse resolution case.
The more localised the distribution is, 
the smaller Doppler velocity we can obtain when degrading the full numerical resolution.
 On top of that,
if we keep the same annular velocity shape but allowing the same amplitude over a broader region that includes the boundary layer,
we would have a strong superposition of Alfv{\'e}n waves with different periods,
leading to very weak signals in a spectrograph.
This is actually suggested in \citet{2018ApJ...856...44A} in order to explain some spicules observations.
However,
neither an adjustment of amplitude nor a more localised Alfv{\'e}n driver model will influence our previous statement that 
the KHI eddies can help to efficiently dissipate the energy in other wave modes.
 
Therefore, 
we can distinguish Alfv{\'e}n modes and kink modes through the properties of fine structures in imaging models and particular shapes of the Doppler velocity and line width properties in spectral models with the original numerical resolution.
Neither small structures nor particular shapes can be observed due to the limitation of the resolving power of real instruments.
So Alfv{\'e}n modes can probably co-exist with kink modes,
leading to enhanced heating,
while being hidden from imaging instruments.
This means that the ubiquitous decayless oscillations in coronal loops can play an important role in coronal heating
by the enhanced heating of unresolved modes.

\section{DISCUSSIONS AND CONCLUSIONS}
\label{sec_conclusion}

In this study, 
we simulated different oscillations in coronal loops,
using a kink driver, 
an Alfv{\'e}n driver and a mixed Alfv{\'e}n and kink driver located at the footpoints of flux tubes.
For all models, 
the oscillations excited in loops can lead to the KHI and generate small eddies.
Especially in the M-model, 
the torsional motions together with transverse motions can help to generate more eddies.
Besides, 
the Alfv{\'e}n oscillations coming from the driver inside the loop and from kink oscillations due to resonant absorption will have phase mixing,
which further enhanced the instability.

We can indeed observe the increase of internal energy and temperature.
The heating is enhanced for the simulation containing the mixed driver,
compared with the other two models.
This means that the mixed modes can lead to a more efficient energy dissipation in the turbulent state of plasma
and that the KHI acts as an agent to dissipate wave energy in other modes.

According to \citet{1983A&A...117..220H}, 
the KHI vortices can also be induced by phase mixed standing Alfv{\'e}n modes.
In turn, 
the small vortices can also reinforce the phase mixing.
This process makes more and more fine structures, 
which can help to dissipate wave energy more efficiently.
However,
in our simulations,
the smaller and smaller scales will become close to the spatial numerical resolution eventually and we can not always observe the finest structures generated in loops.
Generally, if we can capture smaller fine structures, 
the heating effects could be more pronounced.

Forward models can help to compare to the observations.
Fine structures can be observed in the obtained imaging models.
However,
neither Alfv{\'e}n modes nor small structures are observable in the degraded resolution models.
As such, 
the obtained imaging models agree with the decayless oscillations detected through SDO/AIA.
Therefore, 
 this study shows that Alfv{\'e}n waves can probably co-exist with transverse waves in coronal loops, 
 leading to enhanced heating.
 Our spectral models reveal fine structures, 
 the Doppler shift and the line width properties.
 Neither fine structures nor the particular properties can be observed in the coarse resolution models mimicking Hinode/EIS. 
  However,
  beatings can be observed in Doppler velocity and line width in the mixed driver model.
 
 We notice that in the near future,
a new generation of high resolution ground based instruments,
such as Daniel K. Inouye Solar Telescope (DKIST)/Diffraction Limited Near Infrared Spectropolarimeter (DL-NIRSP),
 will help to detect more detailed structures and reveal the energy release processes in the solar corona.
 The potential of this instrument has been recently predicted by \citet{2018ApJ...863..172S}.
 The highest spatial sampling size of the forthcoming DKIST/DL-NIRSP is $0.''03$,
 which is suitable for disk and limb observations,
 while the wide-field mode with a spatial resolution of $0.''464$ will provide coronal observations.
 Within the temperature range in our current models ($\sim {\rm 1MK}$),
 DL-NIRSP may have the ability to recognise the fine structures demonstrated in our forward models due to the high resolution.
 Similarly, 
 forward modeling for next generation instrumentation targeting the recently proposed MUlti-slit Solar Explorer (MUSE) has been done in \citet{2017ApJ...836..219A}. 
 It is shown that most of the features from the TWIKH rolls in coronal loops can be detected with a spatial resolution of $0.''33$ and a spectral resolution of $25 {\rm km~s^{-1}}$. 
 Therefore,
 the future high resolution instruments may help to reveal the turbulent motions in coronal loops and distinguish different numerical models.
 
 We assumed a uniform temperature distribution in the whole simulation domain,
 which can help to recognise the heating effects from waves more clearly.
 According to \citet{2017A&A...604A.130K},
 the mixing between the colder loop and the hotter corona caused a drop larger than 1.5\% in the averaged temperature,
 while simulations with a uniform temperature lead to a rise of about 0.25\%.
 This means that the gradient of the temperature can largely hide the expected heating from waves.
 Once introducing such a temperature gradient,
 we can hardly expect a noticeable temperature increase as in our results here,
 even considering the stronger plasma driving in the M-model.
The larger magnetic field (50${\mathrm G}$) in our models leads to a direct consequence of a smaller transverse oscillation period ($87{\rm s}$).
However, this value is still in the scatter range of relatively shorter loop observations reported by \citet{2015A&A...583A.136A} and \citet{2016A&A...585A.137G}.
Meanwhile, the kink speed in our loops is $c_k\approx 3452 {\rm km~s^{-1}}$,
which is close to the fitting value of $3300 {\rm km~s^{-1}}$ in \citet{2016A&A...585A.137G}. 
  
 Our models still lack realistic solar atmospheric conditions,
 such as vertical stratification due to gravity.
 The vertical non-uniform layer may lead to the reflection of waves,
 which can probably influence the energy carrying capability of waves and the generation of KHI.
 The different transverse distributions of parameters, 
 which are usually studied analytically involving different wave modes \citep{2014ApJ...781..111S,2016SoPh..291..877G},
 will influence the resonant absorption and also the phase mixing of Alfv{\'e}n modes at different magnetic surfaces,
 thereby influencing the dissipation efficiency.
 On top of that,
 the magnetic field variation with height as well as the loop curvature \citep{2004A&A...424.1065V,2009SSRv..149..299V} are also neglected in our current models.
 To clarify their influence on wave heating,
 we will conduct a series of studies on more realistic curved loops in non-uniform force-free magnetic field in future works. 
 
\acknowledgments
{This project has received funding from the European Research Council (ERC) under the European Union's Horizon
2020 research and innovation programme (grant agreement No.724326 and No.647214). 
B.L. is supported by the National Natural Science Foundation of China (41674172, 41474149,  and 11761141002).
M.G. acknowledges the funding from the China Scholarship Council (CSC) and GOA-2015-014 (KU Leuven). 
T.V.D. is supported by the IAP P7/08 CHARM (Belspo) and the GOA-2015-014 (KU Leuven).  
P.A. acknowledges funding from his STFC Ernest Rutherford Fellowship (No. ST/R004285/1).}

\clearpage

\bibliographystyle{apj}
\bibliography{heating_mixed_driver}

\clearpage

\clearpage					

\clearpage					

\clearpage					

\clearpage					

\clearpage					

\clearpage					

\clearpage					

\end{document}